\documentclass[useAMS,usenatbib,usegraphicx]{mn2e}
\usepackage{times}
\usepackage{enumerate}

\title[The STAGES view of red spirals and dusty red galaxies]{The STAGES view of red spirals and dusty red galaxies: Mass-dependent quenching of star-formation in cluster infall}

\author[C. Wolf et al.]{Christian Wolf$^1$, 
Alfonso Arag\'on-Salamanca$^2$,
Michael Balogh$^{3}$,
Marco Barden$^4$,\newauthor
Eric F. Bell$^5$,
Meghan E. Gray$^2$,
Chien Y. Peng$^{6,7}$,
David Bacon$^8$,
Fabio D. Barazza$^{9}$,\newauthor
Asmus B\"ohm$^{10}$,
John A.R. Caldwell$^{11}$,
Anna Gallazzi$^5$,
Boris H\"au\ss ler$^2$,\newauthor
Catherine Heymans$^{12,13}$,
Knud Jahnke$^{5}$,
Shardha Jogee$^{14}$,
Eelco van Kampen$^4$,\newauthor
Kyle Lane$^2$,
Daniel H. McIntosh$^{15,16}$,
Klaus Meisenheimer$^{5}$,
Casey Papovich$^{17}$,\newauthor
Sebastian F. S\'anchez$^{18}$,
Andy Taylor$^{13}$,
Lutz Wisotzki$^{10}$,
Xianzhong Zheng$^{19}$\\
\\
$^1$Department of Astrophysics, Denys Wilkinson Building, University of
  Oxford, Keble Road, Oxford, OX1 3RH, UK (email: cwolf@astro.ox.ac.uk).\\
$^2$School of Physics and Astronomy, The University of Nottingham,
  University Park, Nottingham NG7 2RD, UK.\\
$^{3}$Department of Physics and Astronomy, University Of Waterloo, Waterloo,
  Ontario, N2L 3G1, Canada.\\
$^4$Institute for Astro- and Particle Physics, University of
  Innsbruck, Technikerstr. 25/8, A-6020 Innsbruck, Austria. \\
$^5$Max-Planck-Institut f\"{u}r Astronomie, K\"{o}nigstuhl 17, D-69117,
  Heidelberg, Germany.\\
$^6$NRC Herzberg Institute of Astrophysics, 5071 West Saanich Road,
  Victoria, V9E 2E7, Canada.\\
$^7$Space Telescope Science Institute, 3700 San Martin Drive, Baltimore, MD
  21218, USA.\\
$^8$Institute of Cosmology and Gravitation, University of Portsmouth,
  Hampshire Terrace, Portsmouth, PO1 2EG, UK. \\
$^9$Laboratoire d'Astrophysique, \'Ecole Polytechnique F\'ed\'erale de
  Lausanne (EPFL), Observatoire de Sauverny, CH-1290 Versoix, Switzerland\\
$^{10}$Astrophysikalisches Insitut Potsdam, An der Sternwarte 16, D-14482
  Potsdam, Germany.\\
$^{11}$University of Texas, McDonald Observatory, Fort Davis, TX 79734, USA.\\
$^{12}$Department of Physics and Astronomy, University of British Columbia, 6224
  Agricultural Road, Vancouver, V6T 1Z1, Canada.\\
$^{13}$The Scottish Universities Physics Alliance (SUPA), Institute for
  Astronomy, University of Edinburgh, Blackford Hill, Edinburgh, EH9 3HJ, UK.\\
$^{14}$Department of Astronomy, University of Texas at Austin, 1 University
    Station, C1400 Austin, TX 78712-0259, USA.\\
$^{15}$Department of Astronomy, University of Massachusetts, 710 North
  Pleasant Street, Amherst, MA 01003, USA.\\
$^{16}$Department of Physics, 5110 Rockhill Road, University of Missouri-Kansas City
  Kansas City, MO 64110, USA.\\
$^{17}$Department of Physics, Texas A\&M University, College Station, TX 77843, USA.\\
$^{18}$Centro Hispano Aleman de Calar Alto, C/Jesus Durban Remon 2-2, E-04004
    Almeria, Spain.\\
$^{19}$Purple Mountain Observatory, National Astronomical Observatories,
Chinese Academy of Sciences, Nanjing 210008, PR China.}

\begin{document}
\date{submitted \today}
\maketitle

\begin{abstract}
We investigate the properties of optically passive spirals and dusty red galaxies in the A901/2 cluster complex at redshift $\sim 0.17$ using restframe near-UV-optical SEDs, 24$\mu$m IR data and HST morphologies from the STAGES dataset. The cluster sample is based on COMBO-17 redshifts with an rms precision of $\sigma_{cz}\approx 2000$~km/sec. We find that 'dusty red galaxies' and 'optically passive spirals' in A901/2 are largely the same phenomenon, and that they form stars at a substantial rate, which is only $4\times$ lower than that in blue spirals at fixed mass. This star formation is more obscured than in blue galaxies and its optical signatures are weak. They appear predominantly in the stellar mass range of $\log M_*/M_\odot=[10,11]$ where they constitute over half of the star-forming galaxies in the cluster; they are thus a vital ingredient for understanding the overall picture of star formation quenching in clusters. We find that the mean specific SFR of star-forming galaxies in the cluster is clearly lower than in the field, in contrast to the specific SFR properties of blue galaxies alone, which appear similar in cluster and field. Such a rich red spiral population is best explained if quenching is a slow process and morphological transformation is delayed even more. At $\log M_*/M_\odot<10$, such galaxies are rare, suggesting that their quenching is fast and accompanied by morphological change. We note, that edge-on spirals play a minor role; despite being dust-reddened they form only a small fraction of spirals independent of environment. 
\end{abstract}
%
%


\section{Introduction}

We still do not understand the origin of the morphology-density-relation and the star-formation (SF)-density relation. The local morphology-density relation (Dressler 1980) appears to persist out to $z\sim 0.7$ in the field (Guzzo et al 2007) and to $z\sim 1$ in clusters \citep{Po05}, while the SF-density relation appears to invert its slope between high redshift where star formation first commences at highest densities \citep[see][for results at $z\sim 1$]{El07,Co08} and low-redshift (Lewis et al 2002; Gomez et al 2003) where it avoids overdense regions. This evolution could be explained by continuous formation of non-starforming early-type galaxies throughout cosmic history (Faber et al 2007), where the higher-density regions would be leading the process (Smith et al 2005). As a result, we also find a colour-density relation at low redshift, whereby dense regions are inhabited mostly by red galaxies with old stellar populations; \citet{K04} and \citet{Bl05} have shown that the colour-density relation is the primary trend, while superimposed residual trends in morphology are relatively weak; mostly morphology appears to be driven by colour whatever the density.

In clusters, however, the situation is more controversial: The above picture suggests the presence of an age-density relation at fixed morphology: Thomas et al (2005) claim a strong dependence of early-type ages with environmental density at all masses; while no such evidence is seen in the rich local Coma cluster (Mehlert et al 2003; Trager et al 2008), Wolf et al (2007) see an age-density relation at fixed morphology and low luminosity in the young A901/2 cluster complex. \citet{Po08} again do not see an age trend at higher luminosity in $z\approx 0.4-0.8$ clusters and find that the SF-D and morphology-density relation are equivalent and thus equally fundamental. But if there is no age-density relation at fixed morphology, either close synchronisation between SF quenching and morphological transformation is required, or all transformation producing the patterns must have happened in the distant past such that memories of time scales are wiped out. 

While star formation rate (SFR) is a unique number, morphology is a complex multi-parametric variable: an important distinction must be made between the mass concentration of a galaxy, which is driven by its dynamical history, and its clumpiness, which is driven by star formation and correlates with colour: Over three decades ago {\it anemic spirals} with smooth arms and no signs of clumpy star formation were discovered \citep{vdB76}, which could provide a missing link by being a phase in the transition from a blue star-forming field spiral to a red, non-starforming cluster S0. Also, Poggianti et al. (1999, 2004) and Goto et al. (2003) found spirals without optical signs of star formation in outskirts of clusters, and dubbed them 'optically-passive spirals'. Using an SED-based approach, Wolf, Gray \& Meisenheimer (2005, hereafter WGM05) have described a population of 'dusty red galaxies' dominating the outskirts of the clusters A901/2. Detailed SEDs from the COMBO-17 survey measure dust reddening independent of age-induced reddening. Their sample has red colours due to both dust and intermediate age rather than old age as for regular early-types. Ground-based COMBO-17 morphologies show them to be mostly early spirals \citep{L07}. At least four interesting questions arise here: 

(1) What are the properties of these dusty red spirals? Both passive spirals and the WGM05 dusty red galaxies show no strong optical signs of star formation. The presence of dust and absence of significant optical emission lines, however, motivates a study of obscured star formation with far-infrared data in this paper.

(2) Where are the cluster galaxies undergoing change, such as a decline in star formation? Truncated star-formation disks are common in Virgo spirals and mark a slowly transforming population \citep{KK04}. But most studies of large data sets find that the average star formation rate in star-forming galaxies does not change with environment; instead it appears as if only the fraction of star-forming galaxies declines towards higher densities \citep{Ba04a,VZG08,Po08}. In this paper, we argue that the role of star formation in red spirals has often been ignored \citep[see also][]{H08}, while it is an important aspect in clusters: In fact, we find most star-forming galaxies in A901/2 to be red, while a blue galaxy sample restricts attention to only the most intense tail of star formation.

(3) Are they credible progenitors for cluster S0s? It has been argued that disk fading in field spirals does not produce S0s of the highest known luminosities \citep{Bur05} and correct distribution of bulge/disk ratios \citep{Dr80}, so that the progenitors of bright cluster S0s are still at large. Instead, bulge enhancement \citep{CZ04} or mergers among infalling galaxies \citep{Mo06} may be required. There is indeed evidence for centrally concentrated star formation in cluster spirals, both at low \citep{MW00, KK04} and intermediate redshift \citep{BMA07} that will increase concentration over time. Recent work on higher-quality data also suggests that the concentration differences between spirals and S0s are not as large as once claimed \citep{Lau07,Wz08}.

(4) Which processes initiate the quenching of star formation and the morphological transformation? Both internal and environmental factors may be involved. Internal factors include feedback from supernovae and AGN \citep{Be03,Cr06,Sch07,Tr07} as well as primordial differences in the formation conditions of massive galaxies at high redshift propagating into their present-day behaviour. They are expected to depend on galaxy mass, and upper galaxy mass limits for star formation may evolve with redshift \citep{Bu06}. Environmental mechanisms include ram-pressure stripping by hot ICM (Gunn \& Gott 1972); star formation enhanced by major mergers or tidal excitation from the cluster potential (Barnes 1992; Bekki 1999); loss of gas by suffocation, harassment or strangulation (Larson et al 1980; Moore et al 1996; Balogh et al 2000). E.g., \citet{Mo07} find gradual star formation decline and red spirals in the outskirts of clusters, but truncated star formation and E/S0 galaxies with smooth profiles in the core regions filled by hot gas, presumably due to ram-pressure stripping. It was also speculated, that gravitational heating inherent in structure growth on a cosmic scale (Khochfar \& Ostriker 2007) may quench cosmic star formation as effectively as ad-hoc AGN feedback in semi-analytic models.

Any progress in identifying the mechanism producing red spirals requires the measurement of masses as well as environmental indicators. A low-redshift study by \citet{vdW08} revealed that concentration correlates with mass, while clumpiness correlates with environment. To what extent are then red spirals a mass-specific or a cluster-specific phenomenon? AGN feedback and environment-driven gas depletion are both expected to suppress star formation, perhaps in different mass regimes and over different timescales. At this point, it is not yet clear what their relative role is in suppressing star formation as galaxies grow and structure is built up. To learn more, we need larger samples of galaxies ideally showing their transformation in a few discernible stages. Also, it is important to address evolution in a mass-resolved picture and not with mass-limited samples, as in the latter a mass mix changing with environment produces purely apparent environmental trends.

The intricate problem described above requires a multi-faceted data set of a region in our Universe where galaxies are presently undergoing major change. We have previously identified the supercluster region Abell 901ab/902 ($z=0.167$) as a complex environment that is particularly rich in 'dusty red galaxies' and suggestive of strong galaxy infall activity. The cluster redshift is ideal for obtaining observations that have both high signal-to-noise ratio and complete wide-field multi-$\lambda$ coverage. Such a data set has been obtained by the STAGES team \citep[][hereafter G08]{G08} and provides a high-quality description of a few environmental components coupled to different physics, as well as masses, SEDs, morphologies and star-formation rates for over 1000 cluster members. This effort started with the COMBO-17 survey delivering 17-filter SEDs and high-fidelity photometric redshifts (rms scatter among 800 brightest members is 0.006, see WGM05). The large HST survey at the core of the STAGES initiative added morphologies with $\sim 300~$pc resolution for all cluster members over a $5\times 5$~Mpc$^2$ ($=0.5\degr \times 0.5\degr$) area and the highest resolution dark matter map of large-scale structure \citep{Hey08}. XMM and GMRT observations reveal intracluster gas and AGN \citep{Gil07}. We found that A901/2 is made up of four distinct cluster cores with different masses and geometries, ICM, dark matter and galaxy properties. 

The role of environment in the STAGES data has already been subject to some studies: \citet{Hei08} find that mergers and interactions are found predominantly outside the cluster cores and can be accounted for by the large galaxy velocity dispersion in the cores and the accretion of groups and field galaxies within coherent flows. \citet{Ga08} find no region of enhancement in overall star formation, whether or not obscured star formation is included. With no clear indication of enhanced mergers or triggered star formation, the main impact of environment appears to be a suppression in star formation as already seen in red spirals.

In this paper, we focus on the rich sample of optically passive, red spiral galaxies, which appear to undergo transformation as we observe them, possibly descending from blue spirals and evolving into S0 galaxies. While their SFR is lower than in blue galaxies by a sufficient degree to give them red colours, we choose to relabel them 'semi-passive' given their persistent levels of star formation; as our Spitzer/MIPS data show, these galaxies still harbour significant star formation, which is more obscured than in the average cluster population.  We present the STAGES data relevant to this subject in Sect.~2, and define our environment measures. In Sect.~3 we present relations between IR and optical properties, confirming earlier inferences from COMBO-17 SEDs. We demonstrate the differences between splitting galaxies by blue vs. red and by star-forming vs. non-star-forming. Sect.~4 presents our morphological results and the correspondence of 'red spirals' with 'dusty red galaxies'. In Sect.~5, we discuss galaxy mass functions by type, motivating a mass-resolved investigation of any environmental trends. Sect.~6 shows galaxy properties and type fractions by mass and environment, and finally we discuss a possible evolutionary sequence involving red spirals in Sect.~7.

\section{The STAGES data set}

This study is entirely based on a data set assembled by the STAGES project and recently published in catalogue form (G08). It encompasses high-fidelity photometric redshifts and observed-/rest-frame SEDs from COMBO-17 (Wolf et al 2003), imaging with the Advanced Camera for Surveys (ACS) on board of HST and Spitzer/MIPS IR data, all overlapping on an area of almost $0.5\degr \times 0.5\degr$. Further data have been obtained with XMM, GMRT and GALEX, but are neither part of the recent data release nor used in this study. We select cluster members on the basis of their photometric redshifts. Throughout the paper, we use cosmological parameters of $(\Omega_{\rm m},\Omega_\Lambda,H_0)=(0.3,0.7,70$~km/s/Mpc$)$.

\subsection{COMBO-17 SEDs and photometric redshifts}

The COMBO-17 survey was observed with the Wide Field Imager (WFI) at the MPG/ESO 2.2m-telescope on La Silla, Chile. It used five broad-band filters UBVRI and 12 medium-band filters with resolution $\lambda /\Delta \lambda \approx 20\ldots 30$, altogether covering wavelengths from 350 to 930~nm, to define detailed optical SEDs for objects of $R\la 24$. One of its four fields observed from 1999 to 2001 covers an area of $32\arcmin \times 31\arcmin$ around the A901/2 clusters; see G08 for all technical details of object search, photometry, classification, redshifts and further derived parameters. 

The literature contains several COMBO-17 studies of galaxy evolution using samples defined by photometric redshift. All these restrict themselves to galaxies brighter than $R=24$ to make sure only reliable redshifts are being used. The G08 catalogue of the A901 field contains $\sim 13500$ galaxies with redshift estimates at $R<24$. However, photometric redshift (photo-z) accuracy is a function of photometric noise and declines with fainter magnitudes and field contamination in a photo-z-selected cluster sample increases directly with photo-z scatter. G08 present a detailed discussion of the photo-z scatter and of the completeness and contamination in A901 cluster samples as a function of depth. In Sect.~2.5 we specify the galaxy samples and photo-z accuracies drawn for the purposes of this paper; given the low redshift of A901, we can choose a mass-limited sample while including only the brightest objects with the most accurate photo-z's. 

Here, we only repeat numbers characterizing the overall COMBO-17 galaxy sample. Generally, at bright magnitudes of $R\la 21$ the COMBO-17 photo-z's have an rms in $\delta z/(1+z)$ of $<0.01$, which has been confirmed with spectroscopy on all three COMBO-17 fields with a full data set (CDFS, A901, S11). Using a spectroscopy of 404 galaxies including 249 members of the A901 cluster, \citet{W04} have shown that the cluster photo-z's show an rms scatter of $\delta_z/(1+z) \approx 0.006$ corresponding to $\sim 2000$~km/s. $77\%$ of these galaxies have photo-z deviations from the true redshift of $|\delta_z/(1+z)|<0.01$. Only three objects ($<1$\%) are confirmed to deviate by more than 0.04 from the true redshift and are typically galaxies with unusual SEDs due to contributions from an active nucleus. A deeper comparison is currently only available on the CDFS field, where \citet{W08} find an increase of photo-z scatter to $\sim 0.02$ at $R<23$ increasing further to fainter magnitudes. On the A901 field low-resolution prism spectroscopy of a faint sample was obtained with IMACS at Magellan, which will shed light on the error behaviour at fainter magnitudes when analysed.

The detailed observed-frame SEDs and accurate redshifts have allowed the derivation of robust rest-frame magnitudes for some wavebands including Johnson and SDSS filters and a pseudo-continuum filter at 280~nm intended as a proxy for unobscured star formation rate. At the cluster redshift, the WFI filters used for COMBO-17 cover the rest-frame spectrum from 300 to 800~nm.  The COMBO-17 SED fit involves an age parameter and an estimate of the dust reddening $E_{B-V}$. Stellar mass estimates are based on SED fitting as well and assume a Kroupa (1993) IMF, while a Kroupa (2001) or Chabrier (2003) IMF would yield values within $\pm 10$\% of those (see G08 for all details). Note, that a \citet{S55} IMF increases stellar masses by a factor of 1.8.

\subsection{Aperture effects on masses and SEDs}

The COMBO-17 survey mostly targetted galaxies at $z=[0.2,1.2]$ \citep{W03} with much smaller angular extent than the most massive cluster galaxies in A901. Thus its processing was optimised to obtain high signal-to-noise photometry of faint objects from identical physical footprints (after PSF deconvolution) in all filters. These SEDs represent a central Gaussian aperture with $1\farcs5$ FWHM on a hypothetical infinitely sharp image, and are the basis for the estimation of redshifts, mass-to-light (M/L) ratios and stellar masses, as they have been published now for the A901 field by G08. The SED is normalised with the total object photometry in the deep (20~ksec) R-band image, in which the total light is measured with optimal surface brightness sensitivity, but the colours outside the aperture are ignored. For small objects or particularly large objects without colour gradients this has no consequence. But if large size, low concentration and strong colour gradients are combined, then the SEDs constructed from total magnitudes may be different from the SEDs based on aperture magnitudes. Since our stellar mass estimates depend on M/L ratios from these SEDs, aperture-based SEDs that are not representative of the total galaxy will yield a stellar mass estimate that is incorrect. At the cluster distance the aperture has 4.25~kpc FWHM, and is broadly comparable to the footprint of the fiber aperture in SDSS spectroscopic data at $z\approx 0.08$. This implies that any SED types for galaxies that we discuss later in the paper are defined by the stars seen in the central footprint of 4.25~kpc FWHM. The total SEDs could be different if substantial light with a different SED originated outside this aperture, which may change the colour class to which an object is allocated.

We have investigated the total colours of a few dozen selected galaxies in a separate procedure using SExtractor total magnitudes of all broad bands, although for general use, these colours would be too noisy and object matching is too unreliable, especially given the low sensitivity of the U-band imaging. This has revealed, that aperture values are similar to the total ones across a wide parameter space in our sample. Masses of spirals are reliable at $\log M_*/M_\odot <11$ and masses of E/S0 galaxies are reliable everywhere. But there is an issue with cluster spirals estimated currently at $\log M_*/M_\odot >11$: their total observed-frame $B-R$ colours are on average 0.3~mag bluer than the aperture colours, and this translates into their $\log M/L$ being overestimated by $\sim 0.25$~dex. The $\Delta(B-R)$ corrections range from 0 to $-0.65$~mag, and in the most extreme object (42497) the mass is overestimated by 0.5~dex as a result. This is in contrast to S0 galaxies at $\log M_*/M_\odot>11$, whose total $B-R$ is only 0.04~mag bluer than in the aperture. 

For this reason, most of the 16 spirals in the sample with $\log M_*/M_\odot >11$ are truly just below 11. Almost all galaxies in the field or cluster sample that truly have $\log M_*/M_\odot >11$ appear to be old, red and of E or S0 type. In the absence of a consistent fix for the mass values of high-mass spirals we will omit them later when focussing on the properties of spirals and limit ourselves to the mass range of $\log M_*/M_\odot =[9,11]$ then.

\begin{figure*}
\centering
\includegraphics[clip,width=0.9\hsize]{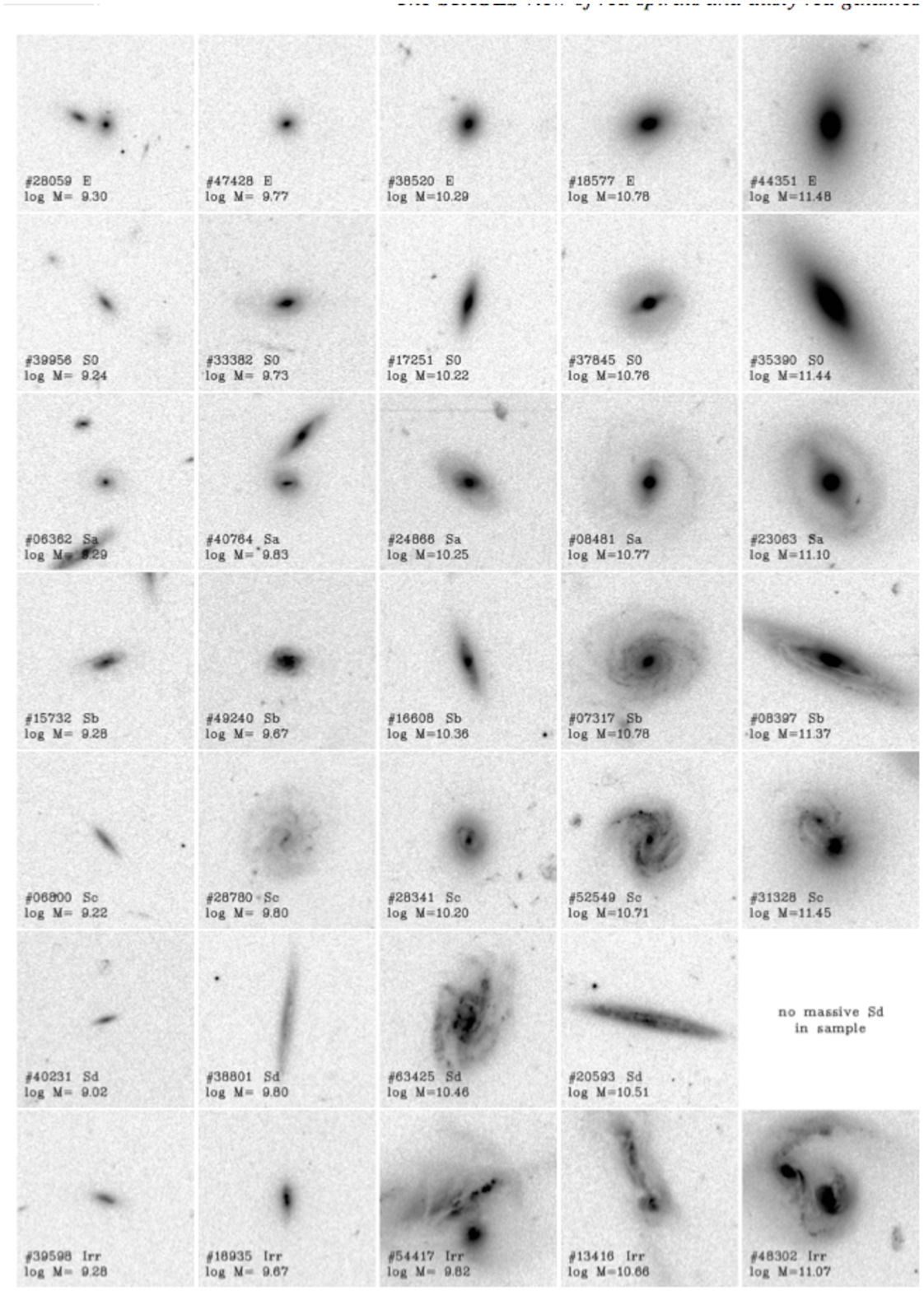}
\caption{Cluster galaxies of different morphological classes in five mass bins across the entire range of $\log M_*/M_\odot =[9,11.5]$ considered in this work (at $\log M_*/M_\odot >11$ masses of spirals are overestimated, see Sect.~2.2). See electronic version for improved contrast of faint features. Image size $12\arcsec \times 12\arcsec$.
\label{cls}}
\end{figure*}

\subsection{HST/ACS imaging and morphologies}

The primary HST observations of the STAGES project cover an 80-tile mosaic with ACS covering an area of $29\farcm 5\times 29\farcm 5$ in the V606W filter during Cycle 13, thus covering $\sim 90$\% of the COMBO-17 area. Various parallel observations of other HST instruments are not part of this study. The mean PSF on these images after data reduction has $0\farcs 11$ FWHM which corresponds to a resolution of 300~pc at the cluster redshift. While the surface-brightness sensitivity of the HST images is lower than that in the 20~ksec R-band image of the ground-based COMBO-17 observations, structural features are seen much more clearly in the ACS images (with few exceptions of LSB tidal arms) due to the $\sim 7\times$ better resolution of HST. 

The published catalogue contains morphological parameters from Sersic profile fitting with GALFIT \citet{Pe02}, but here we have used mostly Hubble types. All galaxies have been visually classified by seven co-authors (AAS, FB, MG, KJ, KL, DM, CW) into Hubble types, including flags on asymmetries and interactions. We distinguish between the types (E,S0,Sa,Sb,Sc,Sd,Irr) and their barred cousins, and define an S0 as a disk galaxy with a visible bulge but no spiral arms. When we notice spiral arms, we put the object into the Sa-Sd range on the basis of our perceived bulge-to-disk ratio, but we suspect that smooth disks and arms as in anemic spirals may bias our judgement towards earlier types \citep{KK98}. The classifiers agree generally well, but small or faint galaxies can be ambiguous when they show little features. For this paper, we use weighted average estimates of the Hubble types, ignoring bars and degrees of asymmetry. We also include all mergers in the irregular class, since they are not of separate interest to this paper (with the exception of Fig.~\ref{rfcols}). Fig.~\ref{cls} shows examples of cluster galaxies from all morphological types we differentiate in this work and covers the entire mass range of our sample. Barred or interacting galaxies are studied in separate works \citep{Ma08,Hei08}.

\begin{table}
\caption{Properties of cluster galaxy sample. \label{sample} }
\begin{tabular}{lrrrr}
\hline  \hline  
$M_*/M_\odot$	range	&  $[9,9.5]$  &$[9.5,10]$ & $[10,11]$  & $[11,12]$ \\ 
\hline 
$R_{\rm tot}$ mean (Vega)	&  $21.22$	&  $20.11$	&  $18.83$	&  $17.23$	\\
$R_{\rm ap}$ mean (Vega)	&  $21.84$	&  $20.89$	&  $19.81$	&  $18.74$	\\
$M_{V,\rm tot}$ mean (Vega)	&  $-18.05$	&  $-19.23$	&  $-20.48$	&  $-22.06$	\\
$z_{\rm phot,min}$	&  $0.122$ & $0.151$ & $0.154$ & $0.154$ \\
$z_{\rm phot,max}$	&  $0.205$ & $0.190$ & $0.187$ & $0.181$ \\
$z_{\rm phot}$ mean&  $0.167$ & $0.171$ & $0.170$ & $0.169$ \\
$z_{\rm phot}$ rms	&  $\pm 0.014$ & $\pm 0.009$ & $\pm 0.006$ & $\pm 0.006$ \\
completeness		&  $>90\%$	&  $>95\%$	&  $>95\%$	&  $>95\%$	\\
contamination		&  $<20\%$	&  $<15\%$	&  $<10\%$	&  $<5\%$		\\
$N_{\rm gal}$		&  $298$		&  $236$		&  $302$		&  $55$		\\
$N_{\rm gal}$ (field) &  $308$		&  $163$		&  $172$		&  $11$		\\
 \hline
\end{tabular}
\end{table}

\subsection{Spitzer/MIPS far-infrared data}

Spitzer has observed a larger area overlapping with $\sim 90$\% of the region covered by COMBO-17 and the HST/ACS imaging, for a total MIPS 24$\mu$m exposure of $\sim 1400$ sec/pixel. The brightest IR source, a Mira star, had to be avoided in the observations, and a large circle of $4\arcmin$ radius around it has been excluded from the analysis. Given the relatively broad PSF with $\sim 6\arcsec$ FWHM, the source list has only been matched to COMBO-17 detected objects. We estimate the sample to be 80\% complete for sources of $97~\mu$Jy, and 70\% of these have been matched to a COMBO-17 counterpart with $R<26$. Conversely, many COMBO-17 sources are not detected at 24$\mu$, and only  $1/3$ of the 600 most massive cluster members are matched to MIPS counterparts. The G08 catalogue contains flux measurements down to the $3\sigma$-detection limit of $58~\mu$Jy, where we estimate a completeness of only 30\%.

Star formation rates (SFR) were estimated for the obscured component using the 24$\mu$m fluxes as a proxy for overall infrared luminosity stemming from dust reprocessing. This component is complemented by UV-based estimates of unobscured star formation mentioned in Sect.~2.1. In the absence of formal errors on the MIPS photometry, we estimate that 24$\mu$m fluxes as well as SFRs are uncertain by up to a factor of two on a galaxy-by-galaxy basis. We also estimate the overall scaling uncertainties of the SFR to remain below a factor of two. Our $3\sigma$-detection limit of $58~\mu$Jy corresponds to an IR-only SFR of 0.14~$M_\odot$/yr at the cluster redshift $z=0.167$ \citep[for more details see][and G08]{B07}.

\subsection{Galaxy samples in this work}

The published catalogue underlying this work contains a flag suggesting a cluster sample defined solely from photometric redshifts. G08 opted for a narrow redshift interval at bright magnitudes taking advantage of the high redshift resolution in COMBO-17, but increase the width of the interval towards fainter levels to accomodate the increase in redshift errors. The selection is defined by including galaxies with $z_{\rm phot}=[0.17-\Delta z,0.17+\Delta z]$, where

\begin{equation}
   \Delta z (R) = \sqrt{0.015^2+0.0096525^2 
                (1+10^{0.6 (R_{\rm tot}-20.5)})  }  ~ .
\end{equation}

This equation defines a half-width that is limited to 0.015 at the bright end and expands as a constant multiple of the estimated photo-z error at the faint end. The floor of the half-width is motivated by including the entire cluster member sample studied by WGM05, and as a result has a selection completeness of nearly 100\% for the brightest galaxies. Overall, this selection achieves high completeness on cluster membership ($>90$\%) while keeping the contamination by field galaxies as low as possible. Increasing redshift errors lead to a dilution in redshift of true cluster members that drives up the fraction of field contamination no matter how narrow a photo-z selection is attempted. It is impossible to refocus a sample once diluted, and one may well choose a wider, more complete sample without much increased contamination. 

We use the G08 cluster definition, and we only use galaxies with reliable photometry, i.e. those having $phot\_flag <8$, and those with morphology data from HST. Finally, we limit our sample by stellar mass. For most purposes we cut at $\log M_*/M_\odot > 9$. This selection contains 891 galaxies, the faintest of which reach $R_{\rm tot}=22$. At the lowest-mass end, photo-z's are spread out over $z_{\rm phot}=[0.122,0.205]$ and we expect a field contamination of up to 20\% (Tab.~\ref{sample} has detailed numbers by mass bin).

For some purposes {in this paper we examine subsamples of galaxies} with $\log M_*/M_\odot >9.5$ or even $>10$; the latter sample contains 357 galaxies that are almost all brighter than $R_{\rm tot}=20$ and within $z_{\rm phot}=[0.154,0.187]$. Of these, 139 sources (39\%) are detected with MIPS at $>58~\mu$Jy. Here, we estimate a field contamination of $<10$\% and a completeness of $>95$\%. We know of a couple of cluster AGN, which are missing from this sample, because their SEDs were affected by blue nuclear light biasing the photo-z estimates (see Gilmour et al 2007). 

\begin{figure}
\centering
\includegraphics[clip,angle=270,width=\hsize]{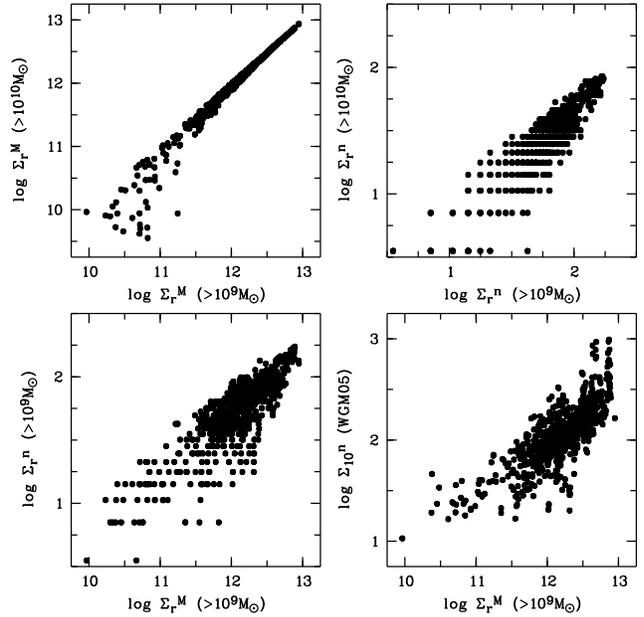}
\caption{Comparison of different environmental density measures in the cluster sample: {\it Top:} stellar mass density within a fixed aperture $\Sigma_r^M$ is not very sensitive to the mass limit of the sample defining the environment, but the galaxy number density $\Sigma_r^n$ is, because low-mass galaxies dominate in numbers but not in mass. {\it Bottom:} the scatter between the robust $\Sigma_r^M$ and $\Sigma_r^n$ is large, whether the aperture is fixed (left, radius $r=300$~kpc) or adaptive as in $\Sigma_{10}$ (right, values taken from WGM05).
\label{compenv}}
\end{figure}

We also use a field sample of 654 galaxies drawn from the same parent data set with a redshift selection that avoids the cluster: we include a lower interval at $z=[0.05,0.14]$ and an upper one at $z=[0.22,0.30]$. These are chosen to prevent even faint cluster galaxies from scattering into them. The upper interval contains much more volume and thus the field sample has a mean redshift of $\langle z\rangle_{\rm field}\approx 0.23$. At masses below 9.5, we estimate the field sample could be 20\% incomplete based on previous experience, but reliable quantification requires spectroscopic confirmation. 

The catalogue published by G08 lists two sets of derived values such as luminosities or masses, one based on the formal photo-z estimate, and another one assuming an a-priori fixed cluster redshift of $z=0.167$, thus preventing the propagation of photo-z errors and peculiar velocities into physical values. Here, we use the fixed-redshift set of values for the cluster sample, but the original estimates for our field comparison sample.

The two different mass estimates for every galaxy also explain why there is indeed no overlap between the field and cluster sample near $z_{\rm phot}\approx 0.13$ as might be wrongly inferred from the above discussion: Only at the faintest magnitudes is this redshift included in the cluster sample, and here the two mass estimates are $\log M_*/M_\odot \approx 9$ in the cluster interpretation ($z=0.167$ forced) and $\log M_*/M_\odot \approx 8.6$ in the field interpretation. As a result of the different masses, these galaxies are included only in the cluster sample and not in the field sample. On the contrary, all field galaxies with $\log M_*/M_\odot >9$ are so bright that the photo-z interval of the cluster begins only above $z=0.14$, so there is again no overlap between the samples.

\begin{figure*}
\centering
\includegraphics[clip,angle=270,width=\hsize]{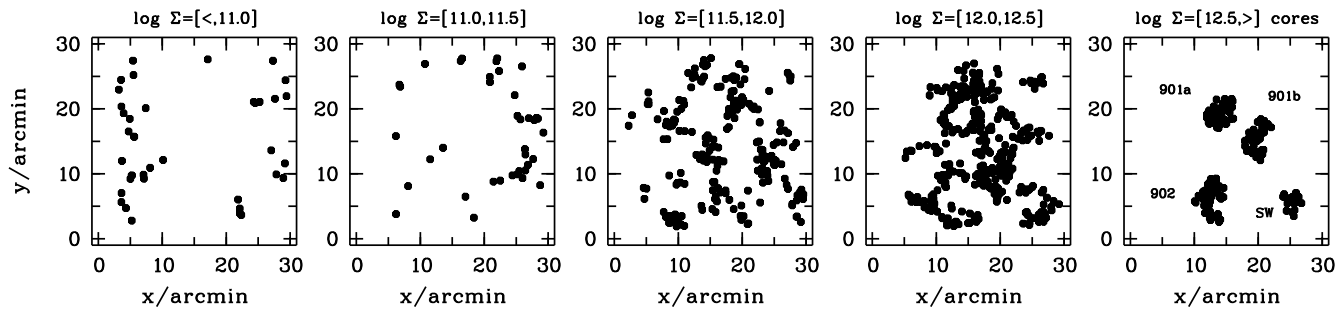}
\caption{Sky maps of the cluster region with member galaxies of $\log M_*/M_\odot > 9$ in five density bins that correspond directly to the five cluster density points used later. The highest-density bin is designated the {\it cores} region in other figures, while the remainder is collectively referred to as {\it infall} region. 
\label{xymaps}}
\end{figure*}

\subsection{Local environment indicators}

The definition of local environment can address different physical components that couple to different physical mechanisms affecting the galaxies: in principle, we need to consider the influence of galaxies at a range of separations because of dissipative encounters and gas heating, the overall dark matter potential because of tidal forces, and the local hot ICM because of ram-pressure stripping.

\citet{L07} have shown that in the A901/2 system trends of galaxy properties with environment are most pronounced when plotted over local galaxy density rather than cluster-centric distance. \citet{Dr80} found the same for an ensemble of clusters. A study by the Galaxy Zoo team, in contrast, finds that the morphology-density relation appears most pronounced when type fractions are plotted against distance to the nearest group centre \citep{Bam08}. Here, we go with our experience on the A901/2 system but explore several measures of galaxy density.

\begin{figure}
\centering
\includegraphics[clip,angle=270,width=\hsize]{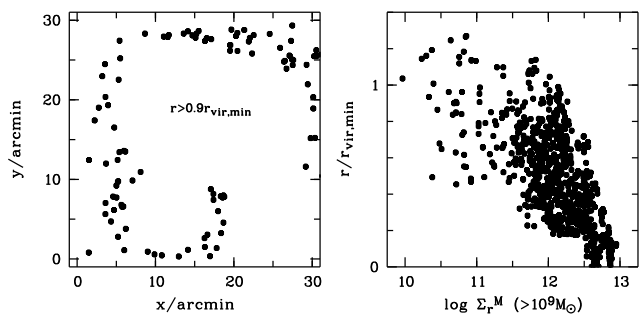}
\caption{{\it Left:} Sky map of cluster member galaxies which are at least 0.9 virial radii from the nearest core in projection. {\it Right:} Comparison of aperture mass density $\Sigma_r^M$ with distance to the nearest core. Virial radii in A901/2 \citep{Hey08} range from 1.1 to 1.7~Mpc . 
\label{rvir}}
\end{figure}

In terms of local galaxy density we compared galaxy number surface densities with galaxy stellar mass surface densities. It turns out that galaxy number densities depend very much on the selection of the underlying sample, which can be by luminosity in different restframe bands, or by stellar mass, and can reach very different depths, simply because faint galaxies dominate in numbers. Conversely, a situation can be imagined where a steep but faint luminosity function (LF) contains the same number of galaxies brighter than some cut-off as a flat but bright LF. The number density would appear the same, although these two are clearly very different environments. In contrast, stellar mass density depends very little on the selection criteria as long as $L^*$ galaxies are included as they dominate the overall mass budget. 

In Fig.~\ref{compenv} we illustrate this point with our cluster sample by measuring four kinds of surface densities inside a fixed aperture with radius $r=300$~kpc $\approx 1\farcm75$: (i) the galaxy number density $\Sigma_r^n$ in units of ${\rm Mpc}^{-2}$, and alternatively (ii) the stellar mass density from all these galaxies $\Sigma_r^M$ in units of $M_\odot {\rm Mpc}^{-2}$; in both cases we include either (a) all galaxies of $\log M_*/M_\odot >9$ or (b) all galaxies with $\log M_*/M_\odot >10$. We compare the two values for mass densities (top left) and for number densities (top right), finding clearly more scatter among number densities. Thus mass density is more robust against variations in the definition of the environment sample. Its insensitivity to the low-mass end will also guard against sample incompleteness at the faint end that may be difficult to quantify.

In the bottom row of Fig.~\ref{compenv} we compare the fixed-aperture mass density against the fixed-aperture number density determined with the same environment selection (left), and also to the variable-aperture number density $\Sigma_{10}$ used by the WGM05 study of this cluster complex (right). The $\Sigma_n$ measures determine the number density within apertures containing $n$ objects, so their apertures shrink in cluster cores and get very large in voids. The result is that $\Sigma_n$ represents potentially different physics in different density regimes. A pair of galaxies separated by $d<r$ will be considered in the same aperture by $\Sigma_r$ irrespective of further neighbours, but appear to be in very different environments assessed by $\Sigma_{10}$, if one of them is closely surrounded by a swarm of Magellanic satellites. In both comparisons we see strong scatter: between $\Sigma_r^M$ and $\Sigma_r^n$ the scatter is a result of comparing a robust measure with a selection-dependent one; between $\Sigma_r^M$ and $\Sigma_{10}$ the scatter is additionally affected by the change of aperture radius and the compactness of substructure clumps. 

In the following, we choose to use the aperture stellar mass surface density  $\log \Sigma^M_{300~\rm kpc} (>10^9 M_\odot)$ as our density measure; Fig.~\ref{xymaps} shows the structure of the A901/2 complex in half decade bins of density. Objects near the edge are eliminated from this plot, where the density aperture was not fully covered by the cluster sample. Still the regions of lowest density are found closest to the edges of the field, which we had placed such that the known cluster cores are arranged around the centre. Higher density structures are progressively closer to the four cores (A901a, A901b, A902, SW group) of the complex, which are depicted very clearly in the map for the highest density bin.

The close proximity of four cluster cores in a small patch of sky is actually a disadvantage for the clean determination of a galaxy's environment, since the infall regions of several cores overlap with other cores. E.g., the centres of A901a and A901b are separated by only $7\farcm35$ ($=1.25$~Mpc) in projection, which is just $\sim 3/4$ of their virial radii, and the two clusters may also interact \citep{Hey08}. The centre of the SW group is $12\farcm2$ ($=2.07$~Mpc) in projection from that of A901b, so their virial radii still intersect in projection; but the SW group is probably more distant than the rest of the system as the mean redshift is higher by $\Delta cz\approx 1500$~km/s. Since the cluster cross-contamination affects the density estimate and the true density mix in an apparent density bin, the true trends will be smeared out.

Our cluster sample focusses on the environment inside the virial radii of the four cores. Few galaxies are more than one projected virial radius away from the nearest cluster core as seen in Fig.~\ref{rvir}. Galaxies in the two lowest-density bins shown in the sky maps of Fig.~\ref{xymaps} have projected distances of only $0.9\pm 0.25$ and $0.75\pm 0.17$ virial radii (mean and rms), respectively. We also see in Fig.~\ref{rvir} that distance to the nearest projected core is not a good estimator of local density in our system, because there is strong scatter between these two quantities.

In the following, we will compare the field to the cluster sample with an emphasis on a well-resolved mass axis: there we simply split the cluster into two parts, referring to the highest-density bin as the {\it cores} sample and to the rest of the cluster combined as the {\it infall} sample. However, we will also use all five cluster density bins when looking at environmental trends with an emphasis on a well-resolved density axis. In these cases, we place a single data point for the field sample at a purely fiducial density($\log \Sigma \approx 10$). We will see later that the properties of the field sample are unlike any of the cluster bins. Irrespective of the true densities at the location of field galaxies, this empirical difference in properties requires to consider field and cluster separately. 

\citet{Ga08} defined a common density estimator for field and cluster in A901 together, given that the 3-D structure of the cluster is unknown. This estimator was based on the photo-z's and their errors and thus led to a strong underestimation of local densities in the cluster: the cluster slice ranges in depth from 120~Mpc for bright galaxies to $>300$~Mpc at the low-mass end, but the cluster system fills only a small part of this pencil beam section volume given a likely cluster extent of only a few Mpc. A pencil beam section in the field sample will have an unknown and different line-of-sight galaxy distribution, such that densities measured there will be on a different scale from those measured in the cluster sample.

\section{Extinction, star formation and colour: how to split the galaxy population?}

In this section, we seek a split of the galaxy population into star-forming and non-star-forming objects, so we can later investigate their properties independently. A common approach has been to fit a colour-magnitude relation (CMR) to the red sequence and place a parallel cut on its blue side to differentiate between red and blue galaxies. In the low-density galaxy field, such a CMR cut works reasonably well for selecting star-forming and non-SF galaxies, although it was established that there is substantial contamination by star-forming galaxies in the red samples \citep{Gia05,Fr07,H08,Ga08}. We point out in the following that such a cut maps less well onto the two desired populations in a cluster such as A901/2. We use optical SEDs and far-infrared properties to explore this issue, and investigate the relation between dust and star formation as seen in the UV and the IR.

\begin{figure}
\centering
\includegraphics[clip,angle=270,width=\hsize]{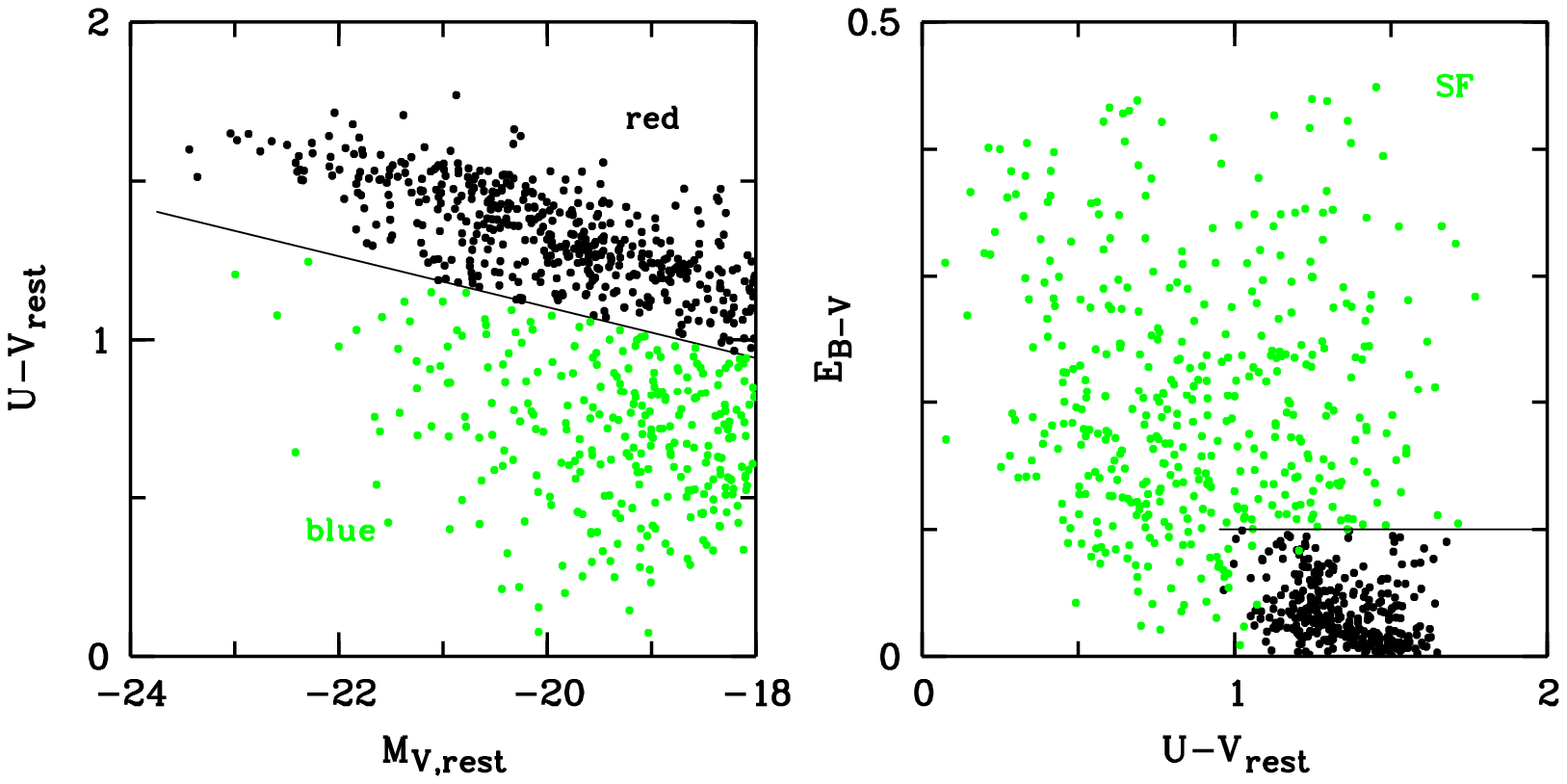}
\includegraphics[clip,angle=270,width=\hsize]{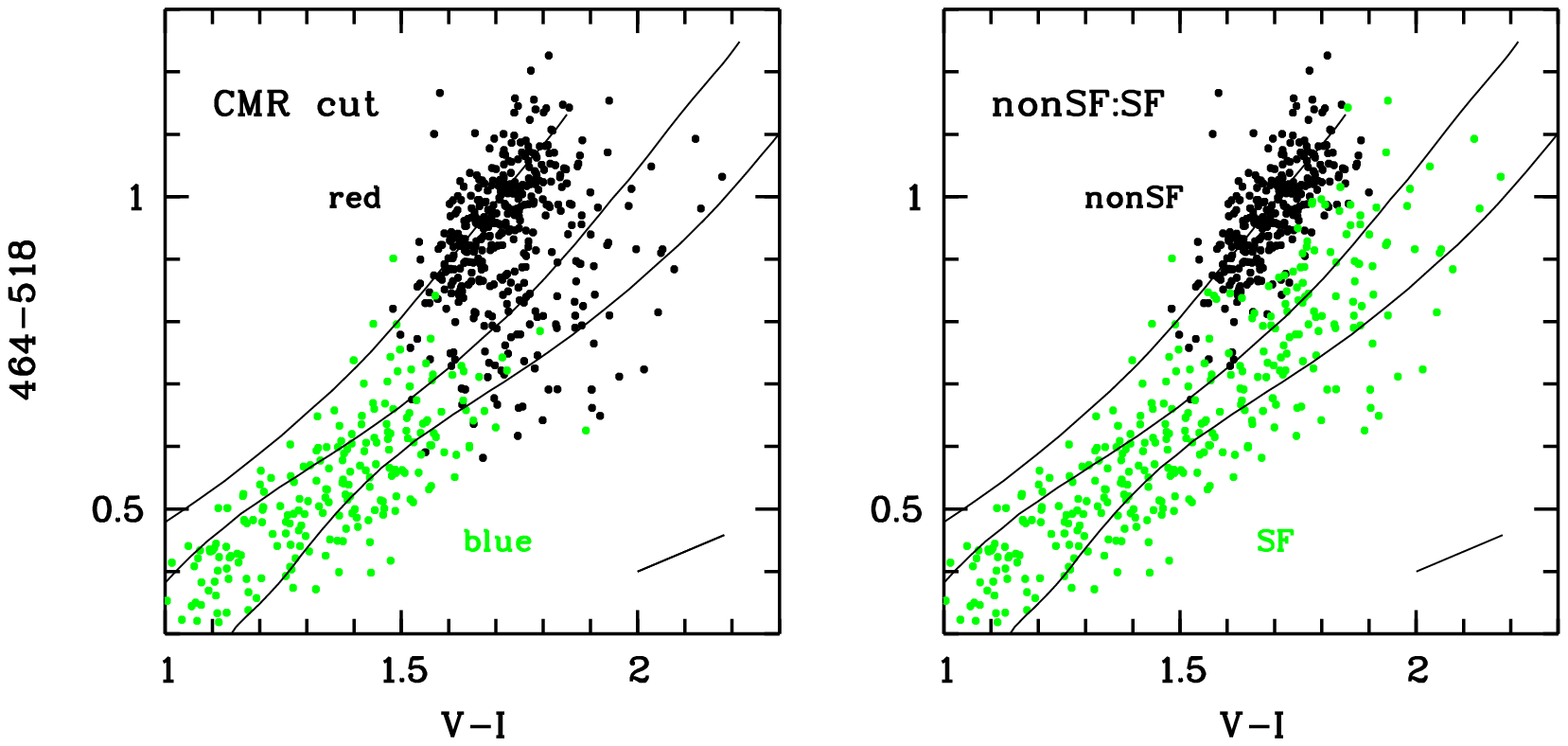}
\caption{Splitting the cluster sample: {\it Top left:} The colour-magnitude diagram of the cluster sample. A standard CMR cut is shown by the line and the red sample is marked with dark points. {\it Top right:} Galaxies in the red cluster sample have a wide range of reddening values. WGM05 found that galaxies on the red sequence are best selected with $E_{B-V}<0.1$, while higher reddening indicates red-sequence contaminants. {\it Bottom row:}
Colour-colour plots with three lines showing model colours for an age sequence with dust reddening of $E_{B-V}=[0.0,0.2,0.4]$; the vertical axis is mostly sensitive to mean stellar age, while the horizontal axis is sensitive to both age and dust reddening (see reddening vector for $\Delta E_{B-V}=0.1$ in bottom right corner). {\it Right:} The cluster population appears to consist of two structures, probably non-star-forming (nonSF) and star-forming (SF) galaxies; these are separated by applying both a CMR and a reddening cut. {\it Left:} Applying only the CMR cut (blue vs. red) fails to discriminate the two structures.
\label{obscols}}
\end{figure}

Since the work of \citet{B04} the COMBO-17 galaxy sample has been split into red and blue galaxies using a cut $0\fm25$ to the blue of the CMR. This cut evolves with redshift as $U-V=1.23-0.40z-0.08(M_{V,h0=0.7}+20.775)$ and its slope is parallel to the CMR slope ($0\fm08$) that was measured in COMBO-17 over the reshift interval $z=[0.2,1.1]$. WGM05 have shown that its extrapolation matches the CMR of the A901 cluster and of the COMBO-17 low-z field. We thus continue to use this cut here and show a colour-magnitude diagram of the cluster in the top right panel of Fig.~\ref{obscols} with the CMR cut as a line.

WGM05 reported red galaxies in A901, which are dust-reddened as much as blue galaxies are, but whose stellar populations are old enough to shift them on top of the red sequence if measured in $U-V$. They still have younger stellar populations than proper passively evolving red-sequence galaxies but together with the dust reddening they end up appearing similarly red. WGM05 dubbed them {\it dusty red} galaxies to differentiate them from passive {\it old red} ones. We will establish here, that these objects actively form stars; any investigation of star formation and its trends with environment should seek to include these objects in the star-forming sample, while studies of the passive population need to exclude them. The trouble with a CMR cut is, that it allocates these red star-forming galaxies to the wrong sample, leading to incompleteness of the star-forming picture in the blue sample, and to contamination by star-formers in the red sample.

The bottom left panel of Fig.~\ref{obscols} illustrates the situation roughly as seen by WGM05: The horizontal axis plots observed-frame $V-I$, which is a colour index formed form two bands redwards of the 4000\AA -break that is affected both by dust reddening and stellar age. Onn the vertical axis, a colour index between two COMBO-17 medium-band filters centred at 464~nm and at 518~nm contains mostly the 4000\AA -break at the cluster redshift; this colour index is strongly affected by age via the strength of the break, but only weakly affected by dust reddening. A short line in the bottom right shows a reddening vector. The long lines represent age sequences of PEGASE models with exponential star-formation histories from very young to 15~Gyr old populations (see Wolf et al. (2004) or WGM05 for more details on the parameters). The top left line is free of dust extinction, while the other two lines are repetitions of this sequence after applying external dust screens with $E_{B-V}$ values of 0.2 and 0.4, respectively. 

When we differentiate the sample by the regular CMR cut (red galaxies are shown as dark points, blue cloud galaxies as light points), we find red galaxies to comprise a focussed dense clump of galaxies consistent with the dust-free models and a further cloud of objects extending to redder $V-I$ colours combined with bluer 4000\AA -breaks. In the model comparison, they reach $E_{B-V}>0.4$ in combination with reduced ages.

The top right panel of Fig.~\ref{obscols} shows again that objects of all restframe $U-V$ colour cover a wide range in $E_{B-V}$ reddening estimates, and illustrates an ad-hoc cut in reddening at $E_{B-V}=0.1$. Red galaxies below this reddening cut were called {\it old red} by WGM05, while red galaxies with higher reddening were called {\it dusty red} by them. We thus apply this further distinction to obtain a cleaned sample of the old red sequence (dark points) of non-starforming galaxies (nonSF). We suggest to consider as a star-forming (SF) sample all remaining galaxies, i.e. the combination of blue cloud and dusty red galaxies and illustrate this split in the bottom right panel. Now, this SF sample appears to be a single structure extending over a range of colours, i.e. over a range of mean stellar ages or (equivalently) specific star formation rates; it also appears to be a separate structure from the concentrated sequence of non-SF galaxies. In summary, we have shown that a suitable colour-colour plot of the A901 cluster shows two structures, which are not well mapped onto a CMR cut, but which can successfully be separated using a combined CMR and reddening cut. We suggest that this cut splits non-star-forming from star-forming galaxies and investigate this claim in the following.

\begin{figure*}
\centering
\includegraphics[clip,angle=270,width=\hsize]{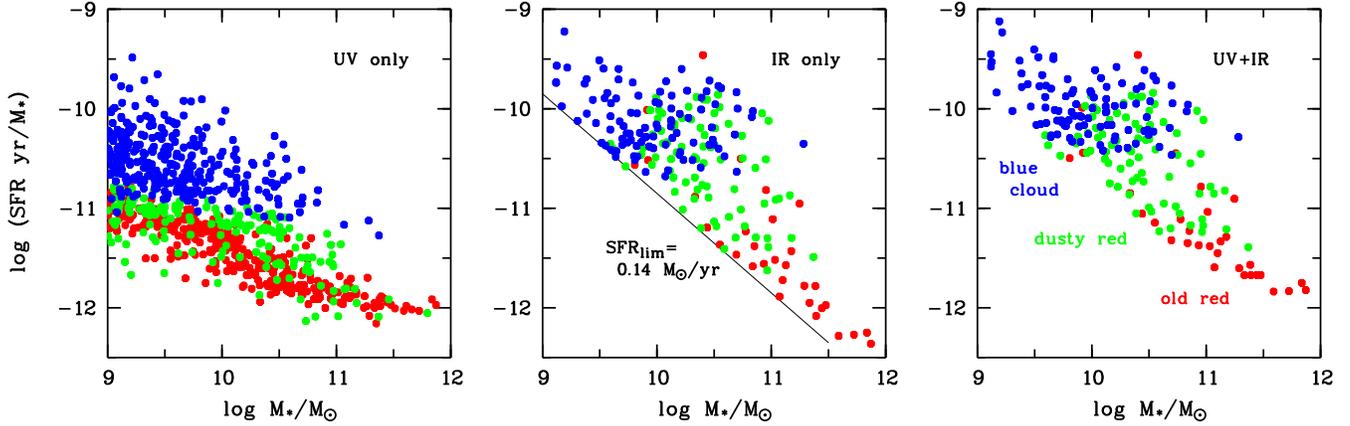}
\caption{Specific star formation and stellar mass: UV-only (left, all galaxies), IR-only (center) and total star formation rates (right, only IR-detections), all split by SED type. The line in the center panel shows the sensitivity limit of the IR data. Most galaxies show more SFR in the IR, and especially dusty red galaxies form stars at a significant rate, $\sim 1/4$ of the rate in blue galaxies at the same mass. Old red galaxies may be affected by UV light from older stars.
\label{sfrm}}
\end{figure*}

\subsection{Are the 'dusty red' galaxies dusty and star-forming?}

If dusty red galaxies really contain dust as suggested by their SEDs and if their younger populations involve ongoing star formation, then we expect to see dust reprocessed UV light in the IR. In the following, we quantify star formation as detected independently in the UV and the IR and estimate dust extinction separately for the star-forming regions and the global stellar population.

From the optical SED we derive the restframe luminosity at 280~nm, which is a proxy for star formation although one that is affected by dust extinction (Kennicutt 1998). From the template SED fit we obtain an estimate of the dust extinction screening the average stellar population. For low extinction levels, the UV luminosity corrected for extinction should be a good predictor of overall star formation. The mean extinction of stars, however, may well be different from that of the very young stellar population and especially from that affecting line-emitting star-forming regions \citep{Ca94}. Also, in the presence of larger amounts of dust, the IR luminosity stemming from the dust-reprocessed light of young stars is preferred as a more direct measure of SFR. Hence, we obtain 24$\mu$m fluxes from Spitzer/MIPS data as a proxy of obscured star formation. A good measure of total star formation can be obtained by summing the UV and the IR contributions. The ratio of IR to UV luminosities ('infrared excess') or derived star-formation rates is then a measure of obscuration in star-forming regions.

There are caveats, however, with both the UV and IR data: light at 280~nm is not exclusively emitted by young stars, but also from intermediate-age or even very old hot horizontal-branch stars. Especially in otherwise dust-poor old stellar populations, it should be a firm upper limit on SFR. Also, IR emission (here at restframe 20$\mu$) can be contaminated by light from a dusty AGN, which we can not identify and remove in this work.

In Fig.~\ref{sfrm} we show the estimates of unobscured SFR from the UV, obscured SFR from the IR as well as the grand total. We estimate $SFR_{\rm UV} = 3.234 \times 10^{-10} \times L_{280}$~$M_\odot$/yr$^{-1}$ and $SFR_{\rm IR}= 9.8 \times 10^{-11} \times L_{\rm IR}$~$M_\odot$/yr$^{-1}$, where $L_{\rm IR}$ is derived from $L_{24\mu }$ using FIR templates; we refer to G08 for in-depth detail. Specific SFR is plotted against stellar mass, and objects are colour-coded by their SED types. The left panel depicts the UV-based estimates, in which the blue cloud and the dusty red galaxies are separated by a fine line corresponding to the usual CMR-parallel cut. While dusty red and old red-sequence galaxies are blended together to some extent, the bluer UV spectrum of dusty red galaxies already increases their formal S-SFR values above the old red sequence. However, as dusty red galaxies are much more extinguished by dust than old red ones, the true extent of their SFR difference is not apparent in the unobscured UV estimates.

In the centre panel, the obscured S-SFR estimate is plotted, but only for the IR-detected sub-sample. The detection limit of $0.14~M_\odot$/yr translates into a diagonal line in the plot; the majority of objects in our mass-selected sample is below this limit and has thus only UV-detections. The detected galaxies, however, show virtually always higher SFRs in the IR than in the UV, including a few formally old red galaxies, where the identification of either UV or IR flux with star formation is often unclear. Finally, the right panel shows the total star formation, which differs little from the IR-only version, because of the general dominance of IR-detected star formation. 

\begin{figure*}
\centering
\hbox{
\includegraphics[clip,angle=270,width=0.25\hsize]{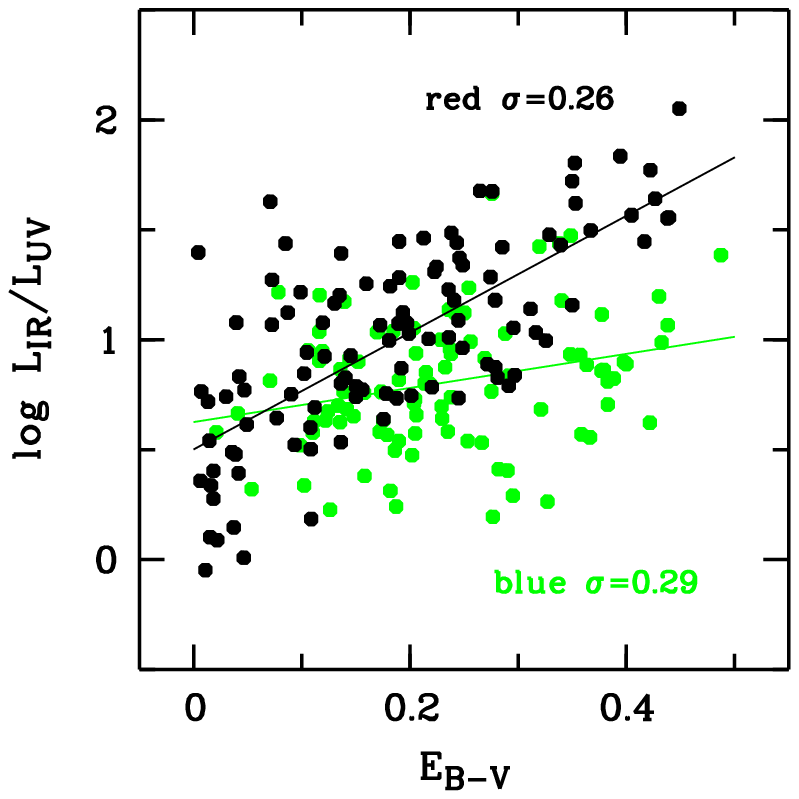}
\includegraphics[clip,angle=270,width=0.75\hsize]{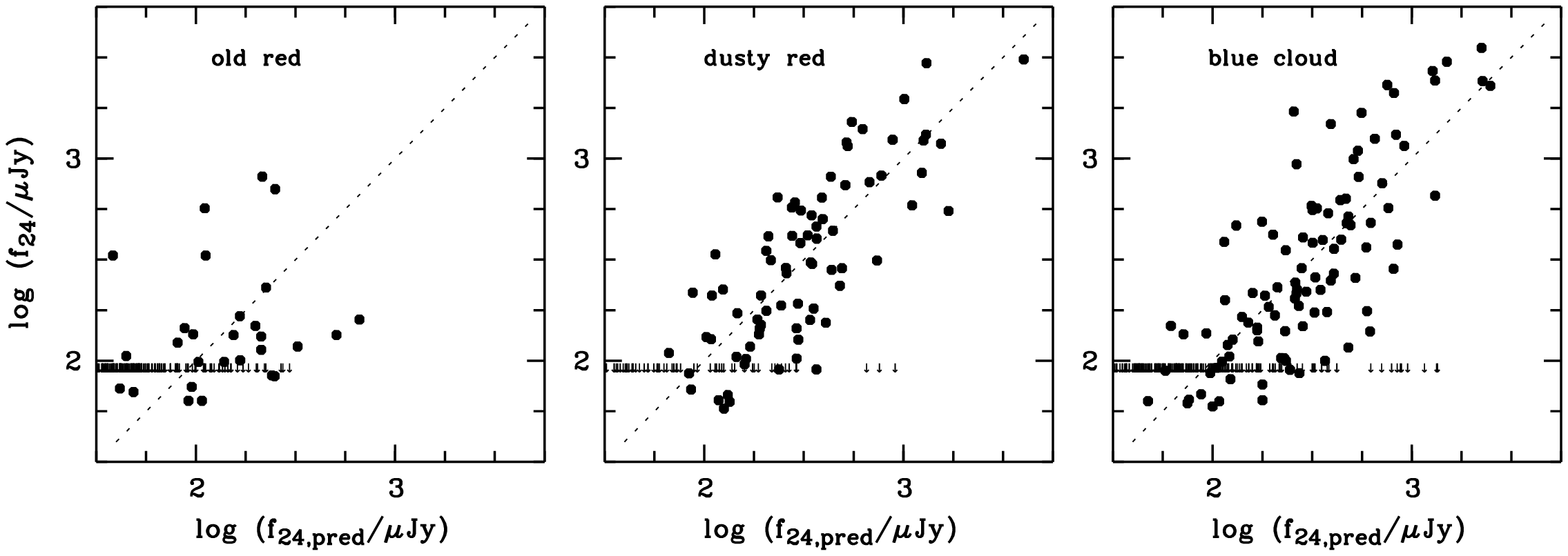}}
\caption{{\it Left:} IR excess vs. optical reddening: the luminosity ratio of IR to UV is correlated with the reddening estimate from the COMBO-17 SED fit for red (dark symbols) and blue galaxies (light symbols). Shown are all cluster members with 24$\mu$m detections and separate fits to dusty red and blue galaxies. {\it Right:} Observed vs. predicted 24$\mu$m flux for all cluster galaxies. Upper limits are plotted at the 80\% completeness limit (see text regarding matching issues).
\label{IRXebmv}}
\end{figure*}

We now compare quantitatively the star formation properties of dusty red galaxies in comparison to blue galaxies, using the mass interval $\log M_*/M_\odot =[10.5,11]$, where 80\% of all blue and dusty red galaxies have 24$\mu$m detections. First, we find that the average S-SFR in dusty red ones is $\sim 3.9\times$ lower than in the blue galaxies. Then we assess the obscuration of star-forming regions as estimated by the average IR/UV ratio in SFR: blue galaxies show $\langle SFR_{\rm IR}/SFR_{\rm UV}\rangle\approx 2.0$ but dusty red ones have an average of $\sim 6.3$ (and extreme ratios of individuals up to 30). Hence, UV-based estimates capture 1/3 of the star formation in blue galaxies, consistent with a large body of literature \citep[see][and references therein]{Cal01}, but only $1/7.3$ in dusty red ones. These values correspond to $A_{280,\rm SF}=1.2$ and $\sim2.15$, respectively, or $E_{B-V,\rm SF}=0.18$ and $\sim 0.32$ if we assume a Milky Way dust law \citep{P92}; at 280~nm LMC or SMC laws would be very similar, and the galaxy mass is certainly typical for the Milky Way. In contrast, the optically estimated $E_{B-V}$ values of the stellar light are similar between dusty red and blue galaxies with mean values of $\sim0.23$ for both. 

We compare the above S-SFR ratio with the ensemble averages of [OII] emission reported by WGM05 as these could be used as an independent albeit less reliable star formation measure: WGM05 found $\sim 4.2\times$ lower equivalent line widths in dusty red vs. blue galaxies; using our U-band continuum luminosities and masses, we derive that at fixed mass the [OII] line luminosities are $\sim 12\times$ lower in dusty red galaxies than in blue ones. We estimate an extinction-corrected [OII]-based SFR using the extinction ratio from the IR/UV continuum luminosity ratio (3 vs. 7.3), and thus find an S-SFR ratio of blue vs. dusty red galaxies of $\sim 5$. The agreement of this number with the S-SFR ratio of $\sim 3.9$ determined above is reassuring and suggests that UV+IR continuum-based SFRs deliver similar results as emission line fluxes if the latter are corrected with the extinction in star-forming regions.

The conclusion is that the dusty red galaxies have on average the largest dust corrections to their SFR values of all galaxies and form very significant amounts of stars indeed. In case of larger obscuration levels the S-SFR of a dusty red galaxy can be almost as high as that in the most star-forming blue ones. On average, however, their total S-SFR at fixed mass is $\sim 4\times$ lower than in the blue cloud. Interestingly, both kinds of galaxies appear to have similar obscuration levels for their overall stellar populations.

Dusty red galaxies should thus be considered siblings of the blue cloud galaxies with reduced star formation; they should {\it not} be considered usual red galaxies undergoing peculiar dusty SF events. The question is then what makes the lower specific star formation rate in dusty red galaxies compared to blue ones? Is it an environmental effect that drives these galaxies through a transition from blue cloud to dusty red, and further to the old red sequence? Or are they simply older than field galaxies at similar mass, because they started forming earlier, and are now optically red enough due to a combination of age and dust to pass a red-sequence cut? Both explanations are plausible and consistent with their known occurrence in cluster regions.

\begin{table}
\caption{Empirical relation between infrared excess and the stellar continuum reddening determined from the COMBO-17 SED fit, using the form $\log L_{\rm IR}/L_{\rm UV} = c + s \times E_{B-V} $. \label{sffits} }
\begin{tabular}{lrrrr}
\hline  \hline  
Sample		& \multicolumn{2}{c}{Dusty red galaxies}	& \multicolumn{2}{c}{Blue cloud galaxies} \\
				&  cluster	& field	&  cluster	& field	 \\
\hline 
slope $s$			&  2.66	&  2.74	&  0.78	&  1.16	\\
error $\sigma_s$	&  0.32	&  0.27	&  0.28	&  0.26 \\
constant $c$		&  0.50	&  0.80	&  0.63	&  0.52	\\
error $\sigma_c$	&  0.08	&  0.27	&  0.07	&  0.09 \\
rms scatter 		&  0.26	&  0.41	&  0.29	&  0.32 \\
$N_{\rm obj}$		&  77		&  26		&  99		&  104 \\
\hline
\end{tabular}
\end{table}

\begin{figure*}
\centering
\includegraphics[clip,angle=270,width=0.85\hsize]{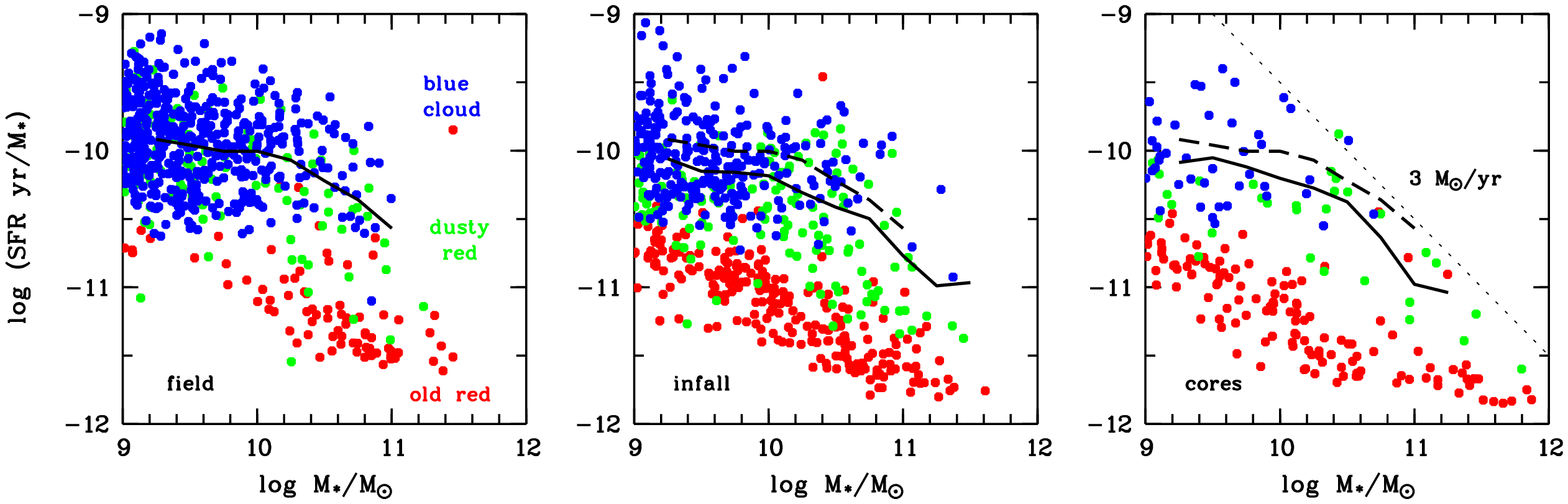}
\includegraphics[clip,angle=270,width=0.85\hsize]{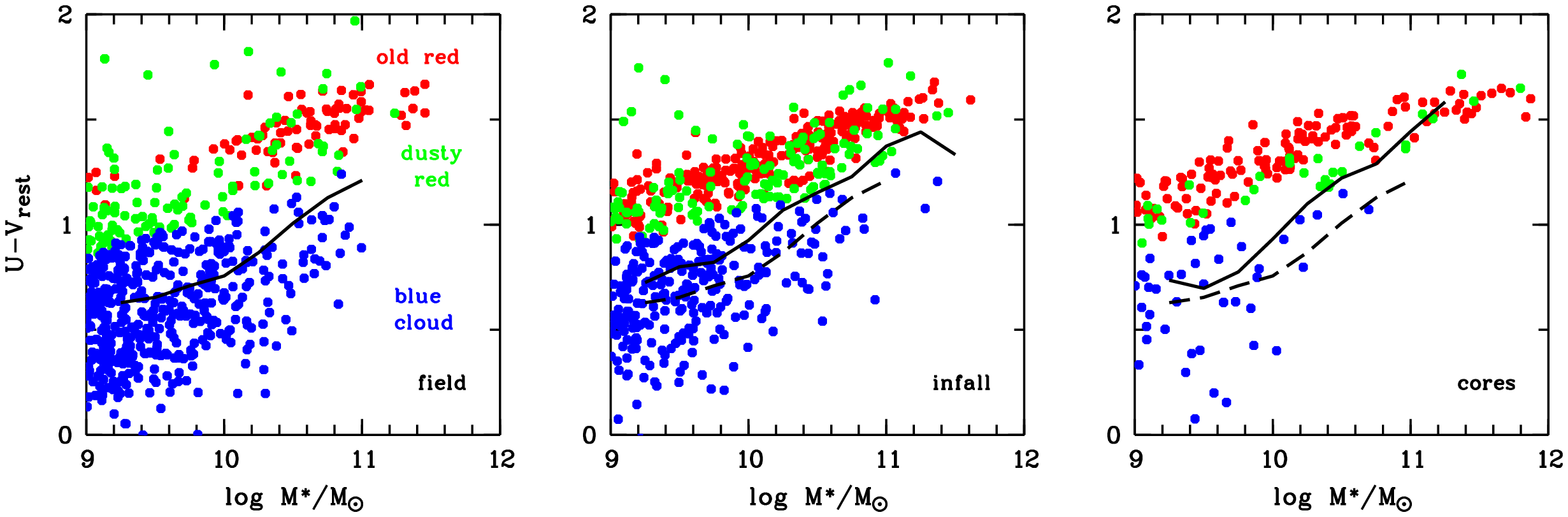}
\includegraphics[clip,angle=270,width=0.85\hsize]{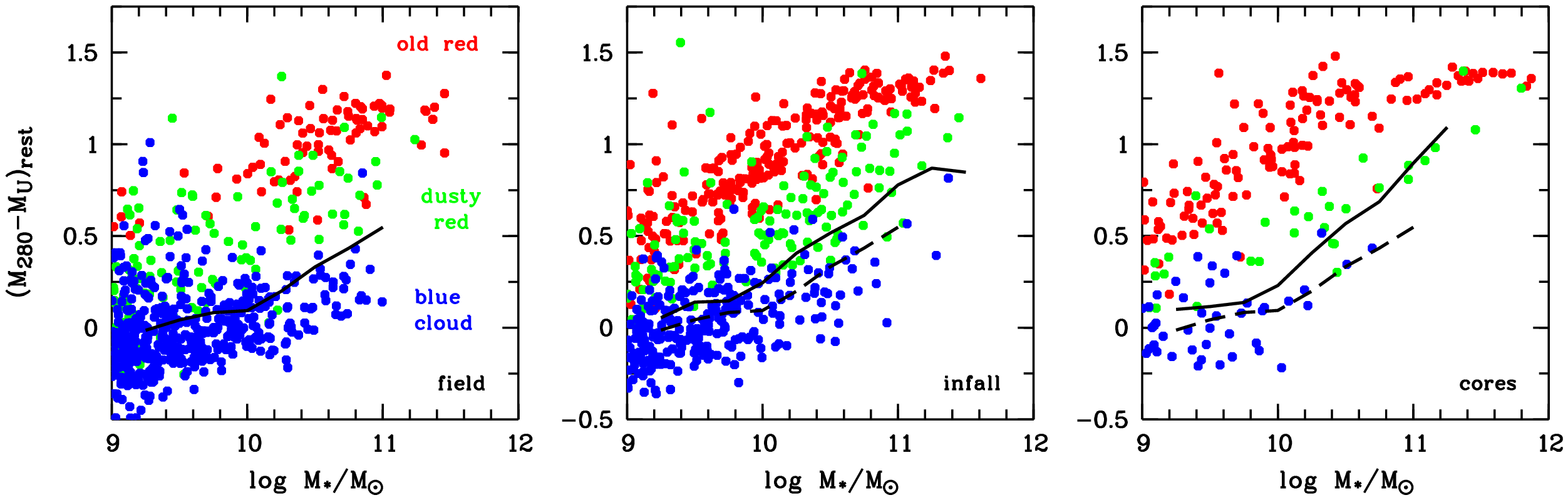}
\caption{
{\it Top:} Star formation-mass diagram of three environments split by SED type. Lines show mean values of star-forming galaxies in half magnitude bins (field dashed for comparison; diagonal line shows constant SFR). {\it Centre:} $U-V$ CMD. Star-forming galaxies and old red sequence are vaguely separated by a green valley in the field, but in the cluster the two blend together and the green valley is populated by galaxies which kept their dust but reduced their S-SFR. {\it Bottom:} $M_{280}-M_U$ CMD. Young stars move the dusty red galaxies away from old red ones to the blue (see Fig.~\ref{xymaps} for definition of infall and cores region).
\label{ssfrM}}
\end{figure*}

\subsection{Can we predict $L_{\rm IR}$ from the optical SED?}

If there is a relation between the extinction of the overall stellar population and that in star-forming regions, then we can predict the star-formation-powered IR luminosity from UV-optical properties. However, we have seen in Sect.~3.1 that dusty red and blue galaxies must have different relations. In Fig.~\ref{IRXebmv} we compare the infrared excess with the $E_{B-V}$ reddening estimate from the optical SED of the stellar continuum. We plot 207 cluster members with 24$\mu$m detections: 108 red galaxies are plotted as dark symbols and comprise 77 dusty red ones ($E_{B-V}>0.1$) and 31 old red ones with lower extinction; the 99 light symbols are blue cloud galaxies. We find correlations that make the UV luminosity in conjunction with the optical reddening a useful proxy of IR flux. It is difficult to predict the 24$\mu$m flux a priori for a number of reasons including dust geometry and temperature. Hence, we just fit a linear relation to the data (see Tab.~\ref{sffits}), separately for red and blue galaxies (any attempt for realistic dust models or more complex relations is beyond the scope of this paper). 

We find a steep slope of $\sim 2.7$ for dusty red galaxies, which also have the higher IR excesses. The rms scatter around the fitted relations is $<0.3$~dex and of the order of our conservative estimate for the IR flux errors; hence, we can not constrain the true scatter of the relations. Blue galaxies show similar scatter but the slope is so low that it is only of limited significance. It also implies the presence of less obscured star formation sites, rendering their star formation a lot more optically visible. An alternative explanation could be that our $E_{B-V}$ values are overestimated for the blue galaxies, being just based on the restricted template set employed in the photometric redshift estimation. We note that the blue sample has many more low-mass galaxies with estimated  statistical reddening errors of up to $\sigma (E_{B-V})\approx 0.2$ as compared to a cluster average of $\sim 0.07$, besides possible systematics of the SED fitting. Dusty red galaxies are mostly at $\log M_*/M_\odot >10$, where errors are smaller. Another complication is that the $E_{B-V}$ estimate and the IR flux are measured in different apertures. 

We now extend our attention to those galaxies not detected in the IR. We predict a 24$\mu$m flux for all cluster members from the UV luminosity and the optically-estimated extinction and compare it with the observed 24$\mu$m flux in the right panels of Fig.~\ref{IRXebmv}. In contrast to the left panel this plot includes 590 IR non-detections as upper limits and further splits the sample by SED type to reveal which types are most affected by non-detections. Overall, low observed 24$\mu$m fluxes coincide with low predictions, but 13 objects ($<2$\% of the sample) deviate by $>2\sigma$ from the relation: they are predicted to show $>500~\mu$Jy, while being apparently undetected at 24$\mu$. A manual inspection of these objects reveals that although they are detected in the 24$\mu$m images, they are not successfully matched to the optical catalogue, because of positional mismatches beyond the chosen tolerance limit. The latter was optimized for reducing wrong identifications among fainter objects but leads to some incompleteness in the matching of brighter sources.

\begin{figure}
\centering
\includegraphics[clip,angle=270,width=\hsize]{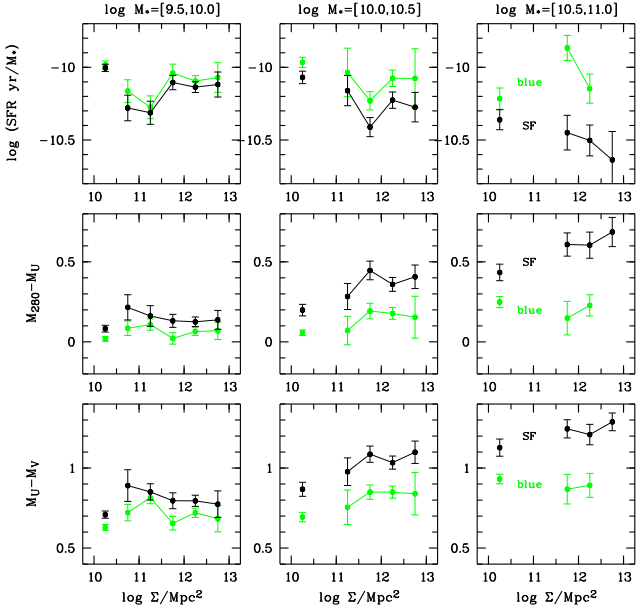}
\caption{Comparing traditional star-forming samples (=blue cloud) with our definition (=SF): Average S-SFR and colour vs. density. At lower mass the cuts are almost equivalent, but at $\log M_*/M_\odot>10$ the blue cloud does not capture all star-forming galaxies. Thus the colour of the blue cloud changes less with density than that of SF galaxies. The latter are redder in denser regions and their S-SFR declines by 40\% to 50\% between field and inner cluster, while the blue cloud alone is consistent with flat S-SFR. The field sample is represented as a single point at a fiducial density of 10.2. 
\label{coltrends}}
\end{figure}

The conclusion is that the IR flux of IR-detected sources in the cluster can be predicted from the optical SED template fit with $<0.3$~dex precision, although there are reasons for not expecting a very strong correlation. There are also no extreme high-obscuration systems in our sample. For our low-redshift field sample, a prediction broadly works as well, with $1.5\times$ the scatter in the cluster. While this statistically describes the bulk of a field sample, we must not extrapolate the relation to the tail of high-obscuration systems expected in the field, in particular at higher redshifts. There, it is a-priori expected not to work, because very obscured stellar populations will not contribute much to the overall galaxy luminosity and SED, and will hence not affect a luminosity-weighted reddening estimate much at all.

The COMBO-17 data are much deeper than the 24$\mu$m obervations, such that sources at the IR detection limit still have restframe UV detections with a median signal-to-noise ratio of $\sim 30$. The most reliable overall star formation indicator would of course be the combination of UV and IR light. But when dust obscuration is weak (which is likely for low-mass galaxies) and no IR detection exists, the most reliable indicator available is an extinction-corrected UV measure; we will use it in the following whenever no IR-detection exists.

\subsection{Star formation and colour split by environment}

Here, we wish to fill in the part of the star formation mass diagram that has not been detected in the IR, in order to assess its appearance despite the caveats of large scatter and uncertain interpretation of the UV light in old red galaxies. Then we compare the results of two different definitions for star-forming galaxy samples: traditionally, star-forming galaxies are taken to be just blue ones, but here we combine blue and dusty red galaxies into the star-formers. We also look at the location of these different SED types in optical/near-UV colour-mass diagrams, and at the overlap between them. 

We begin by plotting specific SFR versus stellar mass, in each of three environments: field, infall region, and cluster cores (Fig.~\ref{ssfrM}, top row). This S-SFR is based on UV+IR estimates when IR flux is detected but on extinction-corrected UV estimates for faint or low-obscuration objects.

In the field, there is a clearly bimodal distribution with star-forming galaxies separated by a valley from the non-SF galaxies. Almost all dusty red galaxies are found among the blue-dominated SF galaxies. This demonstrates that our SED-based distinction suggested in Fig.~\ref{obscols} has succeeded to split the galaxy population into a star-forming vs. a non-star-forming sample much better than a CMR cut would do. This becomes even more important in the cluster environment where red (dusty) star-formers are particularly common while the fraction of star-forming galaxies clearly drops. The mean S-SFR of the SF galaxies declines compared to the field and the valley of reduced S-SFR around $[-11,-10.5]$ is filled in. At this S-SFR virtually all galaxies with detectable dust appear red and at $\log^*/M_\odot >10$ the star-forming galaxies are dominated by the dusty red category. The upper SFR envelope declines with mass so that we find a star formation limit in the cores of $\sim 3~M_\odot$/yr in our A901 snapshot.

\begin{figure*}
\centering
\includegraphics[clip,angle=270,width=0.96\hsize]{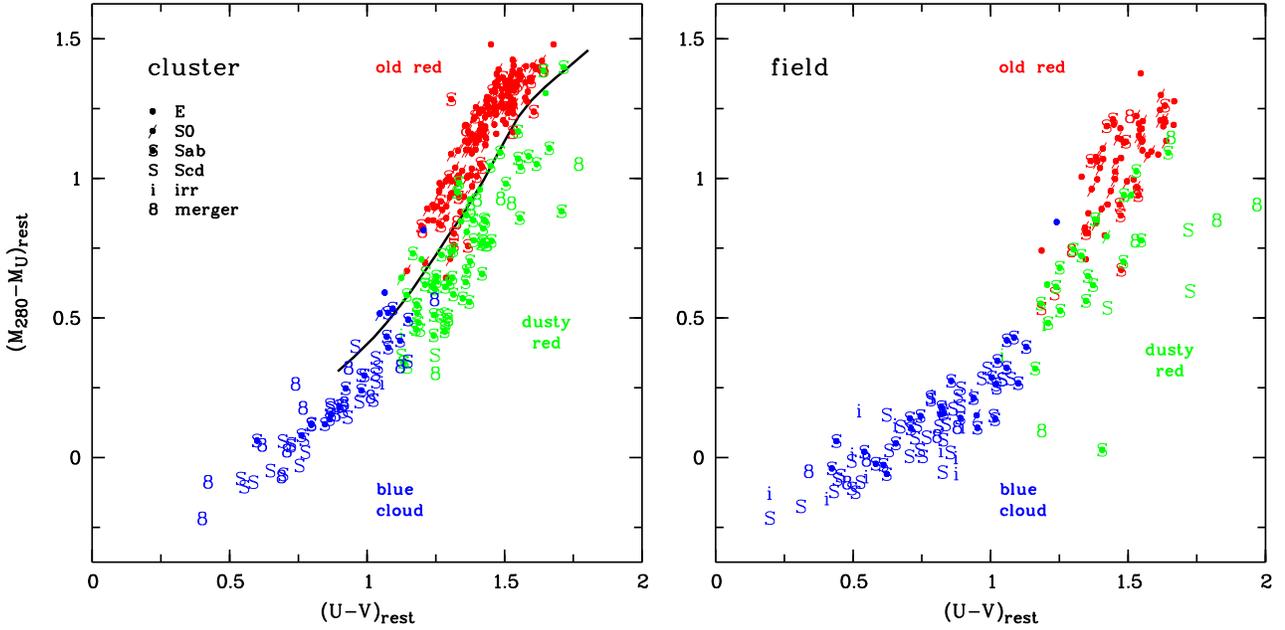}
\caption{Restframe colours of cluster members (left) and field galaxies (right), using only $\log M_*/M_\odot > 10$: symbol indicates morphology and colour resembles SED type. The cluster contains more dusty red than blue galaxies; they are mostly early (Sa/Sb) spirals. At fixed $U-V$ dusty red galaxies are bluer in $M_{280}-M_U$ than old red ones due to young stars. An age sequence with a $E_{B-V}=0.1$ divider between old red and dusty red is shown as a thick line, though $E_{B-V}$ is estimated from the 17-band SED. In the field we see a few outlier objects, very red in $U-V$, which are mostly mergers and edge-on disks. \label{rfcols}}
\end{figure*}

Next we investigate the location of the three SED types in a canonical colour-mass diagram using the canonical restframe $U-V$ colour (Fig.~\ref{ssfrM}, centre row). First we note, that CMR cut separates the blue galaxies from the two kinds of red galaxies by definition. Old red and dusty red galaxies overlap such that a $U-V$ colour-mass diagram provides no means to identify dusty red contaminants to a red sequence sample. Contamination in the red sequence varies from $>80$\% at masses below $\log M_*/M_\odot=10$ in the field to $<20$\% at masses above $\log M_*/M_\odot=11$ anywhere. But freed from the dusty red contamination, the old red sequence in the cluster is a very tight structure in $(U-V)$. We note, that the red envelope of the dusty red galaxies appears to retreat to bluer colours as we go from the field to the cores; as we explain in Sect.~4, this phenomenon is caused by galaxies of very red colour (edge-on spirals and major mergers) becoming rare towards the core.

By combining blue galaxies and dusty red galaxies into what we call the star-forming sample, we also see that the star-forming cloud swipes right across the red sequence and reaches even redder colours than the sequence itself. If the blue cloud alone is taken to represent the star-forming galaxies, its completeness is high in the low-mass field, where it contains $\sim 85$\% of the star-forming galaxies, but in the cluster at $\log M_*/M_\odot>10$ less than half of the SF galaxies are in the blue cloud. Thus, also the mean colours of the SF sample differs from the blue cloud mostly in the higher mass regime of the cluster sample.

However, the overlap between dusty red and old red galaxies discussed above is a consequence of the unfortunate use of the $U-V$ colour index. Instead, a pure near-UV colour such as $M_{280}-M_U$ is more sensitive to young stars, and this we use for the colour-mass diagrams in the bottom row of Fig.~\ref{ssfrM}. Compared to the $U-V$ version, the dusty red galaxies move to the blue, away from old red galaxies, and their overlap is almost removed. 

Finally, we compare the properties of our new star-forming sample ({\it SF}) with those of the blue cloud ({\it blue}) alone by plotting their SFR and colour trends over fine bins in environment in Fig.~\ref{coltrends}. For the blue cloud alone we find only weak or mostly insignificant changes with density, as especially at higher mass the blue sample is very small. At low masses $\log M_*/M_\odot<10$ we find little changes with environment and little difference in the two differently defined samples. In contrast, at masses $>10$ the SF sample shows redder colours and less S-SFR in the cluster as a result of including dusty red galaxies. We find up to a 50\% decline in S-SFR from field to cores in the higher-mass bin. We have thus shown that in a cluster environment and at masses of $\log M_*/M_\odot>10$ the definition of star-forming samples is critical due to the prevalence of red star-forming galaxies that are otherwise often considered an insignificant contribution.

\begin{figure*}
\centering
\includegraphics[clip,width=0.9\hsize]{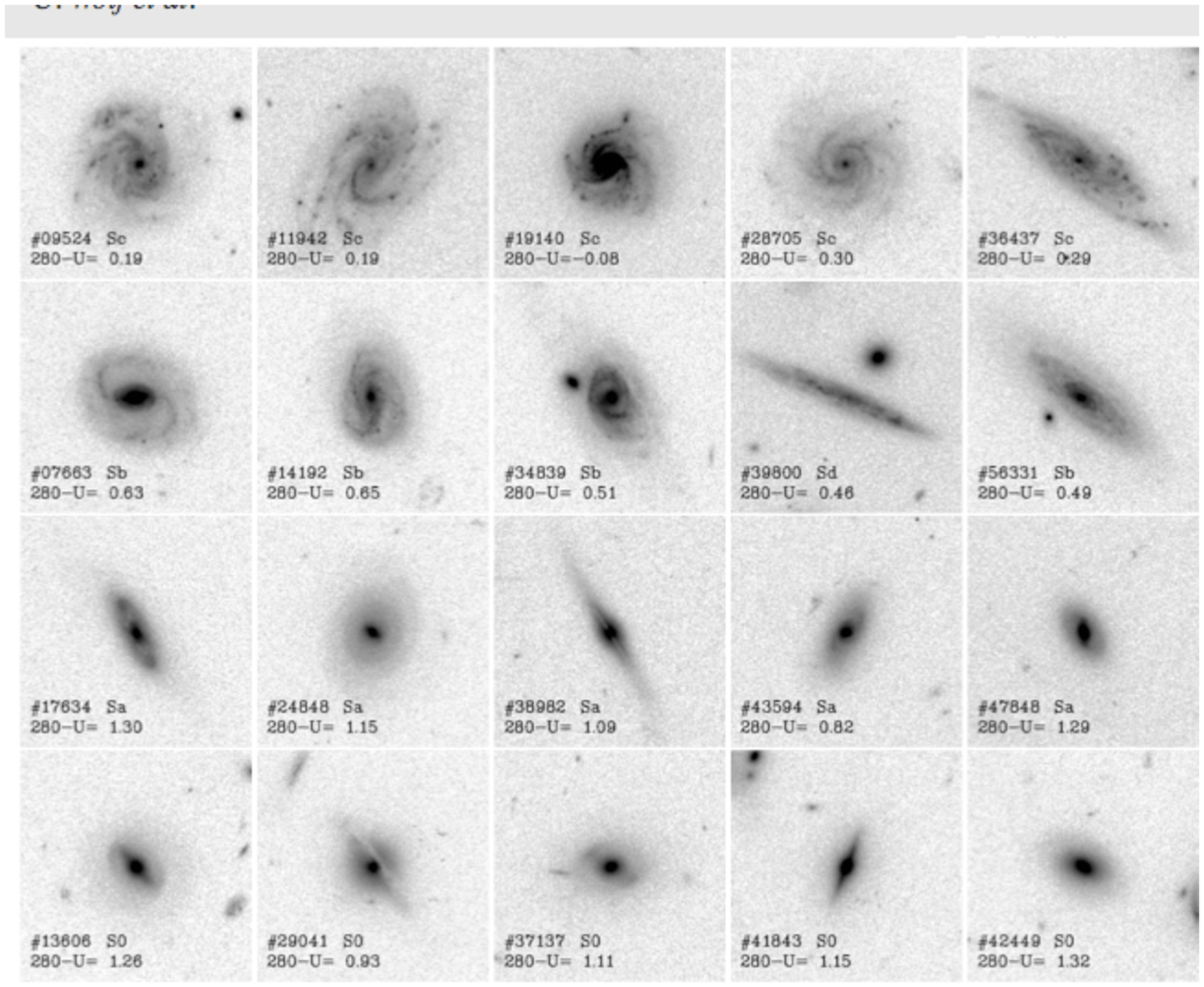}
\caption{A colour-SED-smoothness sequence of disk galaxies ($\log M_*/M_\odot =[10.4,10.6]$). {\it Top:} Blue galaxies with $M_{280}-M_U<0.3$. {\it Second row:} Dusty red galaxies with $M_{280}-M_U=[0.3,0.8]$. {\it Third row:} Dusty red galaxies with $M_{280}-M_U>0.8$. {\it Bottom:} S0 galaxies with $M_{280}-M_U>0.8$ (all old red except for dusty red 29041, which also has a dust lane). We find that our visual classes Sa to Sd are mostly based on bulge prominence, but also on the clarity of the spiral arms, such that dusty red Sa galaxies could also be smoothed spiral versions of bluer "Sb" or "Sc" galaxies. Images are $12\arcsec \times 12\arcsec$.
\label{CSseq}}
\end{figure*}

\section{Morphology and colour: the relation of red spirals and dusty red galaxies}

In this section, we investigate the relation between morphology and colour, paying attention in particular to the morphologies of dusty red galaxies and the subject of red spirals. We use again two restframe colours: The common $U-V$ index straddles the 4000\AA -break and is sensitive to the age of the bulk stellar population, to dust and the presence of young stars; the index $M_{280}-M_U$ shows the near-UV spectral slope and is a lot more sensitive to young stars. We will find that our distinction of star-forming vs. non-star-forming galaxies maps well onto a distinction of spirals/irregulars vs. E/S0s, while this is not achieved by a CMR cut.

We choose to look at the relation between morphology and SED in a UV-optical colour-colour diagram (Fig.~\ref{rfcols}). First, we repeat the relations between SED types encoded in the symbol colour and restframe colours: dusty red galaxies cover a similar range in $U-V$ colours as old red galaxies, but are significantly bluer in $M_{280}-M_U$ at fixed $U-V$ due to young stars. Also, while old red galaxies are clearly detached from the blue cloud and separated by a green valley in the colour diagram, it is again apparent that dusty red galaxies connect smoothly in colour and number density to the blue cloud from which they have been artificially separated by the CMR-cut. This is much more of an issue in the cluster (left panel) than in the field (right), which has few dusty red objects and a more pronounced green valley.

\begin{table}
\caption{Relation between colour and morphology for different definitions of colour types (only for $\log M_*/M_\odot >10$). The WGM05 definition considers $<10$\% of Sb to Irr galaxies as old red, in contrast to a CMR cut that selects 50\% of those in the cluster as red-sequence galaxies. \label{CMcorrel} }
\begin{tabular}{lrrrr}
\hline  \hline  
Hubble type	& \multicolumn{2}{c}{\underline{CMR$-0\fm 25$ cut}}	& \multicolumn{2}{c}{\underline{WGM05-derived}} \\
			&  Red	&  Blue	&  non-SF		&  SF \\ 
\hline 
cluster sample \\
\hline
E		&  73		&  3		&  69		&  7	\\
S0		&  106	&  1		&  96		&  11 \\
Sa		&  50		&  4		&  24		&  30 \\
Sb		&  48		& 25		&  9 		&  64  \\
Sc-Irr	&  15		& 32		&  0		&  47 \\
\hline
total		&  292	& 65		&  198	&  159 \\
\hline
field sample \\
\hline
E		&  29		&  3		&  28		&  4	\\
S0		&  31		&  1		&  27		&  5  \\
Sa		&  19		&  8		&  8		&  19 \\
Sb		&  14		& 25		&  3 		&  36  \\
Sc-Irr	&  8		& 45		&  2		&  51 \\
\hline
total		&  101	& 82		&  68		&  115 \\
\hline
\end{tabular}
\end{table}

Now, we turn our attention to the morphological types that are encoded in the symbols used for the data points: The blue cloud is dominated by Sc/Sd spirals and irregulars, the old red sequence by spheroids (E/S0) and most dusty red galaxies ($\sim 70$\%) are Sa/Sb spirals. A few dusty red galaxies appear to the far right of the main distribution with extremely red colours in $U-V$: these are either merging systems or edge-on galaxies and show prominent dust structures in the ACS images and in either case their SEDs are highly dust-reddened. They are more common in the field than in the cluster, and we already saw in Fig.~\ref{ssfrM}, that extremely red galaxies tend to disappear towards the cluster cores. Considering their morphology this is actually expected, because mergers are more common in the field than in the cluster, and edge-on spirals decline towards the cores alongside the generally declining spiral fraction. In the cores, we then find just a few spirals, which have all reduced star formation and are statistically mostly at moderate inclinations that don't redden their colours much.

The distinction of non-star-forming versus star-forming galaxies (dusty red and blue combined) not only maps well onto the two separate clouds of galaxies in colour space (Fig.~\ref{obscols}), but it also correlates more clearly with morphological types of galaxies as shown in Tab.~\ref{CMcorrel}. A regular CMR cut attributes half of the Sb-Irr cluster galaxies to the red sequence, and in the field still a quarter of them. In the WGM05 system, however, only $<10$\% of Sb-Irr's are considered old red; instead, many more are now classed as dusty red and hence join the star-forming galaxies. The vast majority of (E/S0) spheroidal galaxies are classed as red or old-red/non-SF in either system. We remind the reader, that the definition of the blue cloud is identical in both schemes, and the difference originates solely from the differentiation between old red and dusty red galaxies, and the realization that the latter form an uninterrupted continuation of the blue cloud.

\begin{figure*}
\centering
\includegraphics[clip,angle=270,width=0.76\hsize]{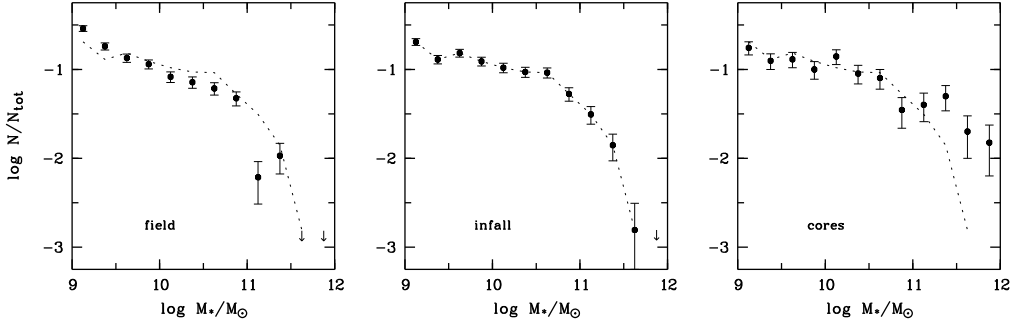}
\caption{Differential mass functions for different environments. The mass function of the {\it infall} region is plotted as a dashed reference line in all panels. Points are normalized to the total number of galaxies with $\log M_*/M_\odot >9$ in each environment. Upper limits are plotted at the location of 1 galaxy in empty bins. Significant differences appear for galaxies with $\log M_*/M_\odot >11$, in which the cores are richer, while the field has a slightly steeper low-mass end.
\label{MF}}
\end{figure*}

\begin{figure*}
\centering
\includegraphics[clip,angle=270,width=0.76\hsize]{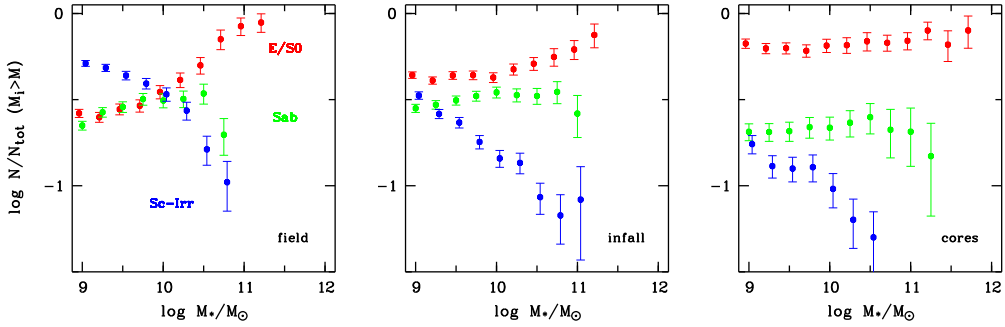}
\caption{Fractions of galaxies with different morphology in mass-limited samples for different choices of the mass cut. Any non-flat trends in these diagrams prove that it is necessary to investigate relations between galaxy properties and environment at fixed mass in order to guard against mistaking truly mass-driven changes for apparently environment-driven ones. Bins with $<3$ galaxies are omitted.
\label{MFcum}}
\end{figure*}

It is thus evident that the red spiral and dusty red galaxy phenomena have strong overlap: the statistics of Tab.~\ref{CMcorrel} show that half of the Sb-Irr galaxies in the cluster are selected by a CMR-type red-sequence cut. Adding in Sa galaxies, we find that $\sim 2/3$ of the A901 cluster spirals are red. In the WGM05 system almost all red spirals belong to the dusty red category ($\sim 75$\% of red Sa/Sb's and 100\% of red Sc to Irr galaxies), and conversely 85\% of all dusty red galaxies are not E/S0 galaxies. Hence, the "dusty red" sample and the "spiral red" sample could be used almost interchangeably; selecting galaxies on the basis of morphology and colour or on the basis of extinction and colour produces similar samples. 

We also note in a non-quantitative fashion, that for a given Hubble type red spirals show less structure within their disks and spiral arms than the blue spirals do, which is a sign of reduced star formation. The extreme cases of disks with no discernible structure are classified as S0 galaxies. Fig.~\ref{CSseq} shows example galaxies of a complete colour sequence, where substructure declines with redder colours, stretching from blue spirals to S0 galaxies.

In summary, red spirals are mostly not on the proper red sequence at all, as opposed to S0s, of which 90\% are there; instead, they just populate the red tail of the blue cloud much more strongly in the cluster than they do in the field. We note that the average morphological type of the cluster galaxies among blue cloud, dusty red and old red samples is Sbc, Sb and S0, respectively ($\log M_*/M_\odot >10$), and the difference between blue Sbc and dusty red Sb might be entirely a result of a smooth arms bias or of disk fading. Thus, blue and dusty red galaxies differ little in overall morphology and dust content, but mostly in colour and SFR.

It is important to avoid a possible fundamental misconception about the origin of red spirals: E.g., \citet{Cas07} attribute the red-sequence contamination to edge-on spirals, which is appropriate in a field sample. However, in our cluster sample red spirals are more common than even blue spirals and Fig.~\ref{CSseq} shows only a few edge-on examples. Furthermore, our measurements show that an average red spiral in the cluster has a similar dust extinction but reduced star formation compared to an average blue spiral; all this argues very strongly against an orientation effect and in favour of a cluster-related origin of our rich red spiral sample (see Sect.~7.1).

\section{Mass functions by environment}

We already know that massive galaxies tend to be red spheroids while low-mass galaxies are more often disk-shaped and blue, independent of environment. A relation between morphology and density, e.g., will thus be biased by a changing mass mix of the population with environment, if drawn from a mass-limited sample. Any trends seen in such relations could potentially be driven only by mass, and are not clearly attributable to the environment. Hence, we briefly investigate to what extent the galaxy mass function in our sample depends on environment.

Fig.~\ref{MF} shows the mass function of the galaxy population in the field, infall and core regions, and each of them is normalised by the total number of galaxies at $\log M_*/M_\odot >9$ in its environment. Significant differences exist at $\log M_*/M_\odot >11$, where the cores are much richer than the infall region, while the field tends to be less rich. The field also appears steeper at the faint end than the mass function in the infall region and cores. The mass function of field galaxies does not look like a single Schechter function, but more like a combination of several mass functions for different galaxy types with different shapes and slopes. Its faint end slope is thus not well-determined in contrast to the flatter mass function in the cores. Formal linear function fits in the mass range $\log M_*/M_\odot =[9,10.5]$ yield $\alpha_{\rm field}=-1.47\pm 0.04$ and $\alpha_{\rm cores}=-1.16\pm 0.07$.

The determination of faint-end slopes in cluster mass functions is traditionally considered challenging due to the effects of completeness and field contamination. \citet{RG08} show that at least for a cluster as nearby as Virgo spectroscopy is required, as many photometric red-sequence members are truly behind the cluster. They find a faint-end slopes of $\alpha=-1.28\pm 0.06$ for Virgo that is consistent with the field slope, while we tend to see a trend with environment among dwarfs (given our redshift and method we should not suffer from background contamination). However, this discussion is probably not settled yet, as there is disagreement about the Virgo mass function: while the seminal work by \citet{SBT85} found $\alpha\approx -1.4$, modern estimates range from $\alpha=-1.28$ as above to $\alpha=-1.6$ for the outer Virgo regions and the low-density galaxy field following \citet{Sab03} who claim that inclusion of LSB galaxies makes a critical difference at lowest luminosities; these could be stripped and faded descendants of small-bulge spirals, see \citet{Bos08}. 

In either case, the clear differences at the high-mass end we see lead to different galaxy type mixes due to intrinsic mass-type relations. This point is further illustrated in Fig.~\ref{MFcum}, where we plot the fractions of galaxies with different morphology in mass-limited samples for different choices of the mass cut (here, error bars are necessarily correlated). Any non-flat trend in these diagrams illustrates how the sample mix changes with mass. As the giant-to-dwarf ratio increases from the field over the infall region to the cluster cores, the type mix of mass-selected samples changes alongside the mass mix. This is a necessary one out of two possible factors explaining the differences in type mixes seen when comparing the environments shown in Fig.~\ref{MFcum}. The other possible factor is an explicit environmental dependence of galaxy properties. A better way of investigating such trends is to look for changes in galaxy properties with density that can be seen in narrow mass ranges, over which the mass composition is unlikely to change significantly with environment; this is the subject of the next section.

\begin{figure}
\centering
\includegraphics[clip,angle=270,width=\hsize]{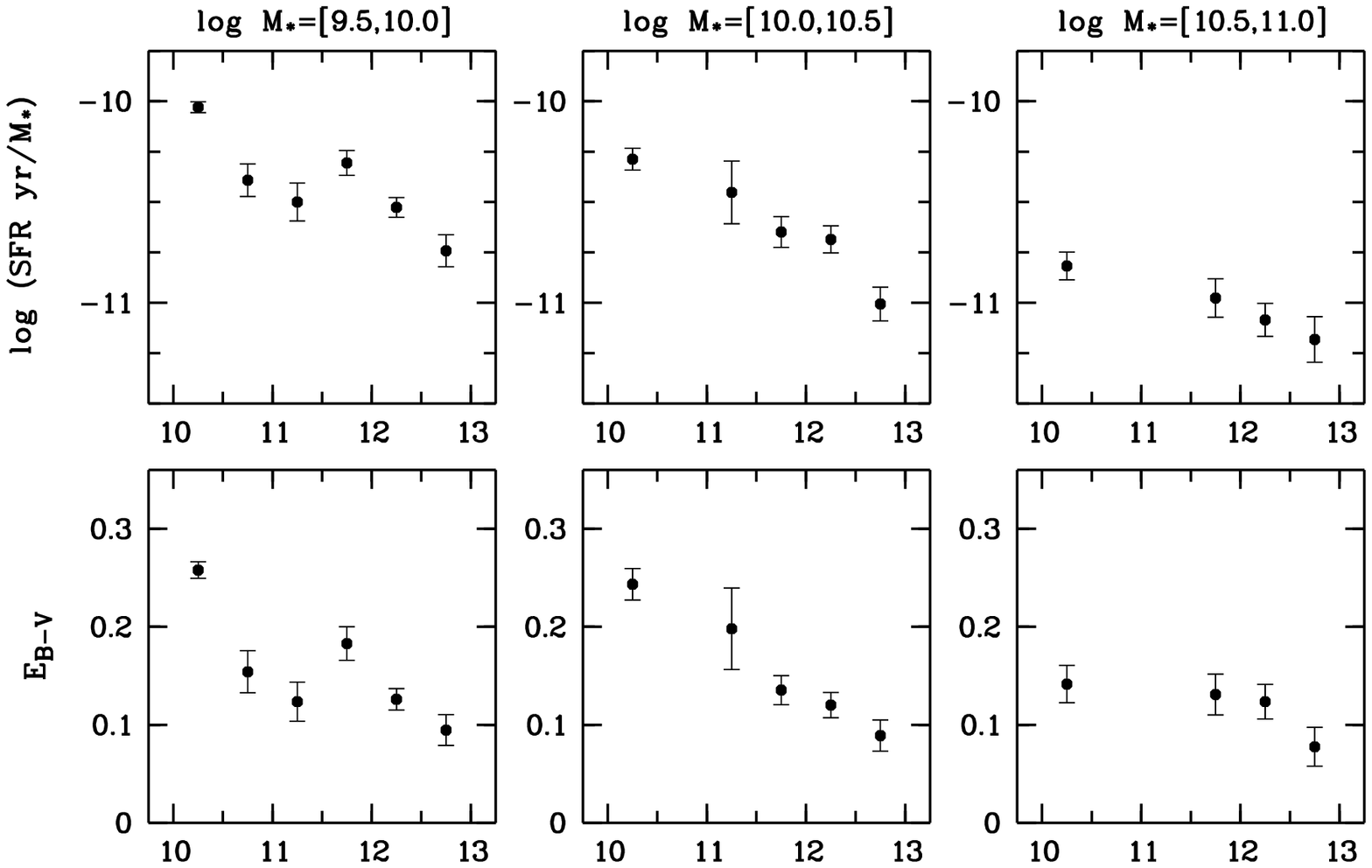}
\includegraphics[clip,angle=270,width=\hsize]{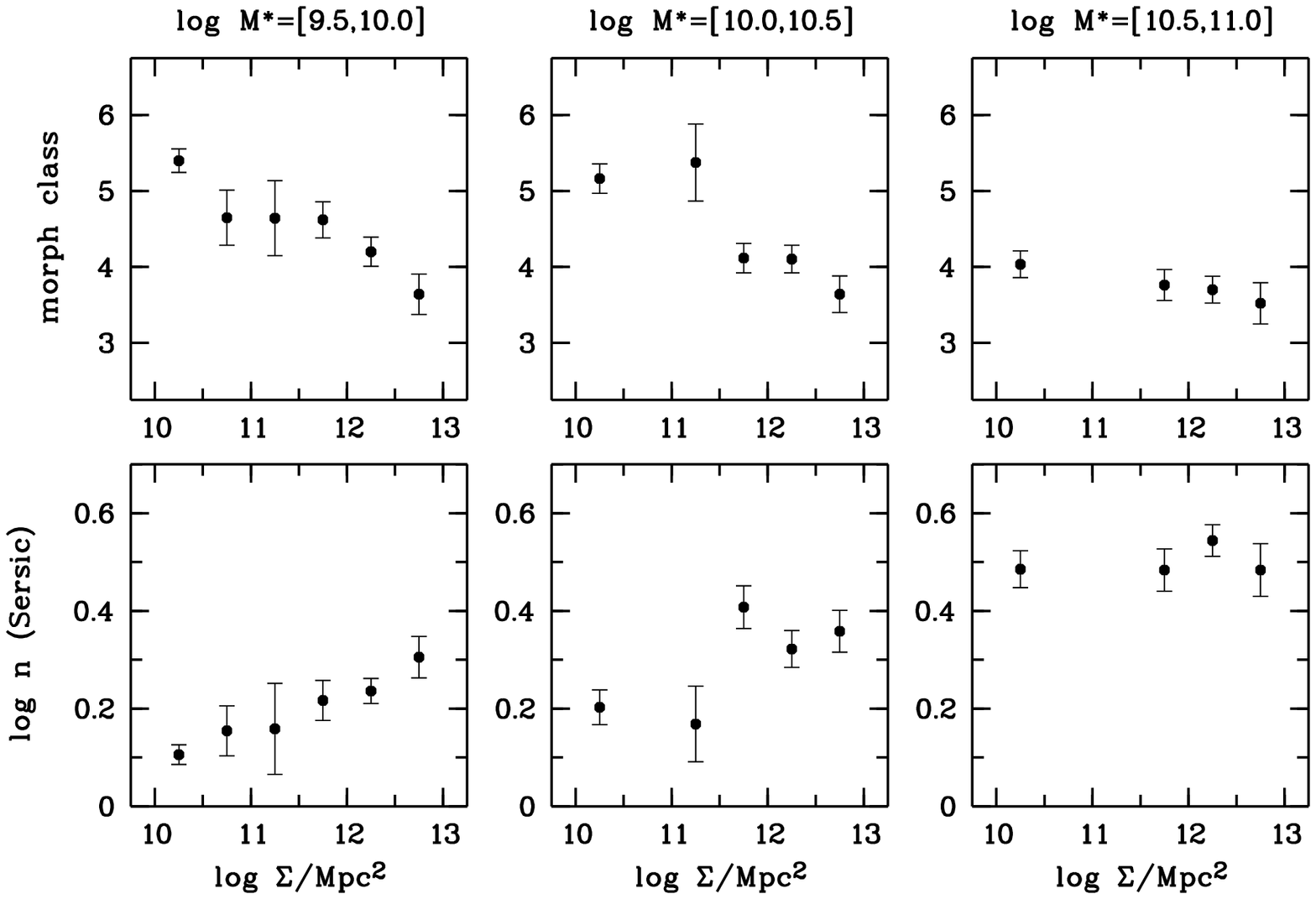}
\caption{Galaxy properties as a function of density: Star formation rate and stellar extinction decline at all masses continuously towards higher density. In contrast, morphological changes appear discontinuous at intermediate mass. Only small changes are observed at $\log M_*/M_\odot >10.5$. The leftmost point represents the field value and cluster points correspond to the density intervals used in Fig.~\ref{xymaps}.
\label{PDrel}}
\end{figure}

\begin{figure}
\centering
\includegraphics[clip,angle=270,width=\hsize]{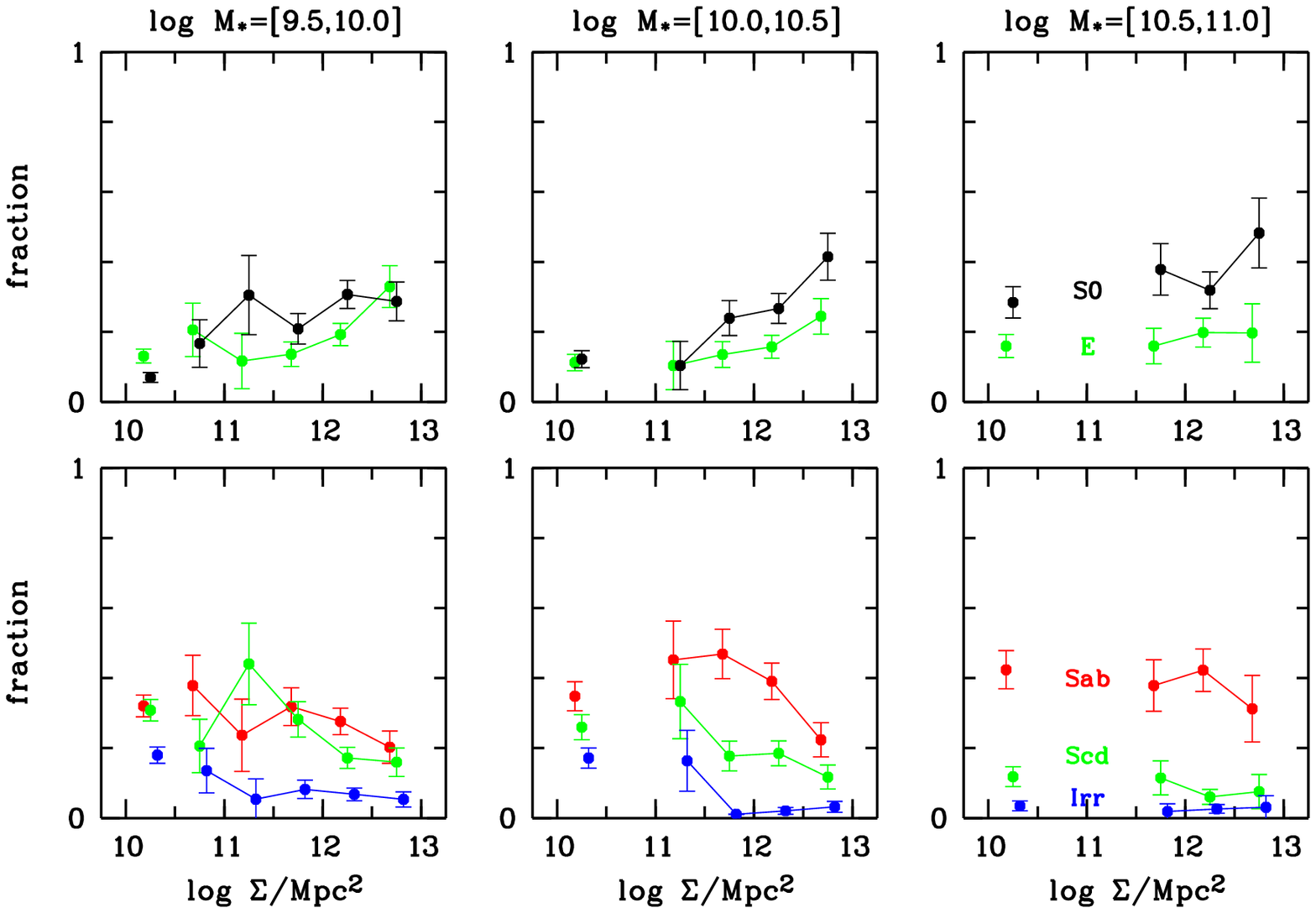}
\includegraphics[clip,angle=270,width=\hsize]{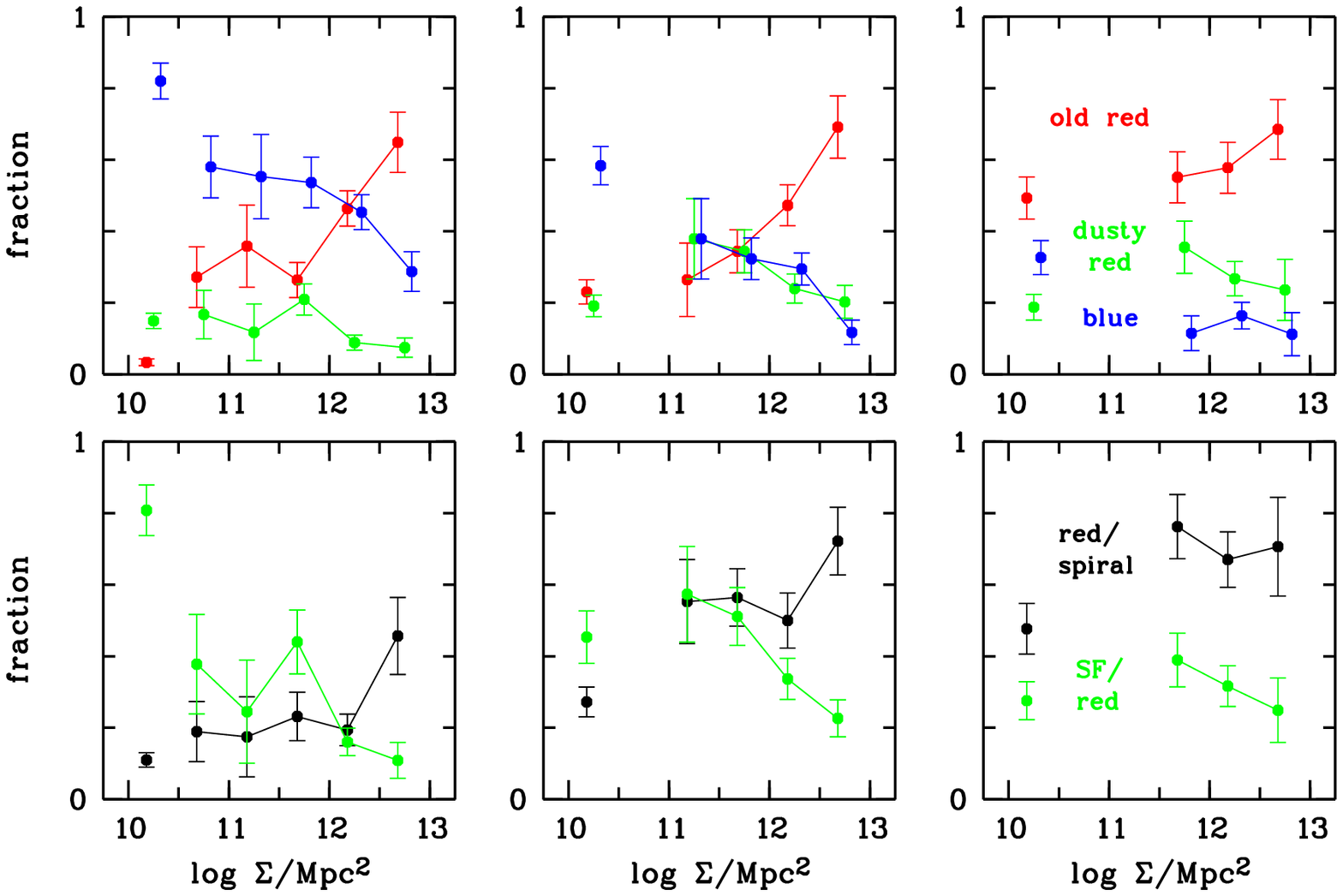}
\caption{Galaxy type fractions as a function of density: Dusty red galaxies (and thus red spirals) are most prominent at intermediate mass and density, and result from colour change due to declining SFR in otherwise blue spirals. Towards the cores they are mostly replaced by S0s. Errors are Bayesian assuming fractions can only vary within $[0,1]$. For different types points are horizontally offset for clarity only.  \label{TDrel}}
\end{figure}

\section{Trends with environment and mass}

\subsection{Trends in star formation and morphology}

In this section we investigate trends of star formation, dust extinction and morphology with environment and how these trends together give rise to a preferred habitat for dusty red galaxies/red spirals. We try to separate trends with environment from trends with mass, as galaxy properties depend explicitly on mass and the mass mixture of a galaxy sample depends explicitly on environment, thus giving rise to spurious trends in mass-integrated samples. Hence, we plot mean properties or type fractions of galaxies over density using well-resolved bins in both mass and density. However, this will dilute our sample and produce larger statistical error bars. We keep a single separate data point for the field sample using a purely fiducial density of $\log \Sigma = 10.2$ (see Sect.~2.5 for a motivation of this choice), and plot cluster points representing exactly the density bins illustrated in Fig.~\ref{xymaps}; these points are centred between their interval limits and slightly offset between types for clarity. Fig.~\ref{PDrel} and \ref{TDrel} share a common structure: Purely environmental trends can be seen within a single panel of the figure, while trends with mass are seen in the comparison of different panels in a row. The mass bins cover half decades from $\log M_*/M_\odot = 9.5$ to $11$, above which very little trends can be seen. Type fractions have Bayesian errors assuming the prior that they are within $[0,1]$. Bins with $n_{\rm all}=1$ or those which are incomplete in a given mass range are omitted.

Fig.~\ref{PDrel} shows trends in mean specific SFR and dust reddening $E_{B-V}$ as well as two morphological measurements: the visually determined mean Hubble type ($[3; 4; 5; 6] = [{\rm S0; Sa; Sb; Sc}]$) and the Sersic index obtained with GALFIT. These diagrams tell us three important things: 

\begin{table*}
\caption{The composition of the four cores in comparison: A901a has the fewest star-forming galaxies (see fraction of blue and of 24$\mu$-detected galaxies); the X-ray luminous A901b has the most S0s and fewest red spirals; A902 has the highest spiral fraction, but may appear at enhanced density due to projection. \label{cores} }
\begin{tabular}{llllllllll}
\hline  \hline  
core		&  $N_{\rm obj}$  & $f_{\rm blue}$ & $f_{\rm 24\mu,M>10}$  & $f_{\rm red~spiral}$  & $f_{\rm giant}$  & $f_{\rm E}$ & $f_{\rm S0}$ & $f_{\rm Sab}$ & $f_{\rm Scd-Irr}$ \\ 
 \hline 
A901a	& 60 & $0.11\pm 0.04$ & $0.29\pm 0.05$ & $0.22\pm 0.05$ & $0.14\pm 0.04$ & $0.40\pm 0.06$ & $0.28\pm 0.05$ & $0.14\pm 0.04$ & $0.19\pm 0.04$ \\
A901b	& 50 & $0.26\pm 0.05$ & $0.43\pm 0.07$ & $0.09\pm 0.04$ & $0.13\pm 0.04$ & $0.32\pm 0.06$ & $0.44\pm 0.07$ & $0.15\pm 0.04$ & $0.11\pm 0.04$ \\
A902	& 70 & $0.27\pm 0.04$ & $0.40\pm 0.05$ & $0.26\pm 0.04$ & $0.09\pm 0.03$ & $0.25\pm 0.04$ & $0.27\pm 0.04$ & $0.30\pm 0.05$ & $0.19\pm 0.04$ \\
SW group	& 20 & $0.26\pm 0.09$ & $0.41\pm 0.10$ & $0.26\pm 0.09$ & $0.26\pm 0.09$ & $0.31\pm 0.10$ & $0.36\pm 0.10$ & $0.26\pm 0.09$ & $0.13\pm 0.07$ \\
 \hline
\end{tabular}
\end{table*}

\begin{itemize}
\item The specific SFR and the dust extinction (row 1 and 2) of the overall stellar population both change gradually with density. It appears as if there is already a change to less star formation and less extinction from the field to our outermost infall points; while the error bars overlap, a formal decline is seen in mass bin. This may correspond to the group pre-processing much discussed in the literature \citep[e.g][]{Wil05}.
\item There is no difference in the morphological properties (row 3 and 4) between field galaxies and those in the outermost infall bin at $\log M_*/M_\odot >10$ and there appears to be a more sudden change further into the infall region. Below 10, however, the morphology appears to change gradually with density again.
\item At $\log M_*/M_\odot >10.5$ all trends appear much weaker, because field galaxies are already mostly of early type and have little star formation. Below 10.5, the overall changes from field to cores are virtually identical between the two mass bins despite possible differences in the infall region.
\end{itemize}

Unfortunately, we could not sensibly plot trends of the mean extinction in the star-forming regions of star-forming galaxies due to the shallow IR sensitivity limits in the field sample. However, in Sect.~3 we have already established broadly that dusty red galaxies dominate the SF sample in the cluster while having little relevance in the field, and have more obscured star formation regions than blue galaxies. 

Fig.~\ref{TDrel} shows trends in several galaxy type fractions with environment. The first two rows plot Hubble type fractions; the third row compares the fractions of the three SED types, blue cloud vs. old red and dusty red galaxies. The last row shows in dark points the fraction of spirals is redwards of the CMR cut (old red plus dusty red); the light points show the fraction of dusty red galaxies out of all red galaxies, which we take to be the fraction of star-forming galaxies among red galaxies (SF/red). We draw the following conclusions:

\begin{itemize}
\item The strongest morphological changes (row 1 and 2) are seen within the infall region of the mass bin $\log M_*/M_\odot =[10,10.5]$ and are associated particularly with the demise of all kinds of spirals and a steep increase in the S0 fraction. Scd and Irr galaxies decline first, while Sa/Sb spirals only decline closer to the cores. At $\log M_*/M_\odot =[10,11]$ S0s dominate over E's everywhere in the cluster, but at higher masses (not shown here) ellipticals are clearly dominant.

\item There is little change in the Hubble type fractions (row 1 and 2) between the field and the lowest-density infall point that is plotted, especially above $\log M_*/M_\odot =10$. 

\item Dusty red galaxies (row 3) are most dominant at intermediate mass and density, but they play virtually no role at $\log M_*/M_\odot <10$; by implication the same applies to red spirals (row 4). At $>10$ the rise of red spirals in the infall region is due to SFR decline that is {\it not} accompanied by morphological change, so that blue field spirals are replaced by red infall spirals; going further towards the cores all sorts of spirals are gradually replaced by red S0s (row 1). 

\item At $\log M_*/M_\odot <10$ in contrast, blue field spirals are replaced by red infall spheroids as a result of colour change being accompanied by synchronised morphological change. Hence, the red fraction among spirals depends strongly on mass and ranges from $\sim 20$\% at low mass ($<10$) to $\sim 80$\% at high mass ($>10.5$). At $\log M_*/M_\odot >10$ it changes strongly from field to the first infall point and stays high throughout infall, while the overall spiral fraction declines.

\item The contamination of star-forming galaxies in the red sequence (SF/red, row 4) is generally low in the cores, but high in the infall regions at intermediate mass due to dusty red galaxies being common there, and high in the field sample at low mass ($<10$), where the CMR cut enters into the contiguous structure otherwise known as blue cloud and leads to 80\% of the red sample being star-forming.

\item Again, at $\log M_*/M_\odot >10.5$ all trends are weaker.
\end{itemize}

With respect to red spirals our particular cluster environment appears quantitatively consistent with the cosmic distribution of red spirals at low redshift as studied very recently in the Galaxy Zoo project: \citet{Bam08} show that at low density roughly 10\% of $\log M_*/M_\odot=[9.5,10]$ spirals are red, which increases to nearly 50\% at $\log M_*/M_\odot=[10.5,11]$; finally, at high density this number converges to $\sim 75$\%. Despite differences in mass determination and density estimator, these numbers match ours surprisingly precisely. Another strong signal that could appear spurious, data-specific or possibly A901-specific, is the extreme ($>80$\%) SF contamination of the red low-mass field sample; again this number matches perfectly the recent results of \citet{H08}. Hence, we also conclude that although A901/2 has been selected by us to be a convenient laboratory harbouring a range of different environments in a diverse cluster complex that is accessible to single-shot observations, it appears to still be a completely representative environment, from which physical conclusions can be generalised.

\subsection{Differences between the four cores}

There are in total exactly 200 galaxies in the four cores of our sample at $\log M_*/M_\odot >9$. Of these, the three cores with Abell numbers (901a, 901b and 902) contain between 50 and 70 each, while the SW group has a less rich dense core with only 20 galaxies (the fact that these galaxy numbers here are integer multiples of 10 is entirely coincidental). The SW group also differs from the other cores in its high giant-to-dwarf ratio: 25\% of its core galaxies have $\log M_*/M_\odot >11$, compared to only $\sim 10$\% in the Abell cores (see Tab.~\ref{cores}).

The Abell 901a core appears most circular in shape and is further characterized by the lowest fraction of blue galaxies, i.e. $\sim 10$\% vs. $\sim 25$\% for the others, and less significantly the lowest fraction of galaxies with $\log M_*/M_\odot>10$ and 24$\mu$-detections, i.e. $\sim 30$\% vs. $\sim 40$\%. It could thus be the most evolved core among the four, although the various possible evolution indicators do not present a clear picture. We consider a regular shape, high E/S0 fraction, low spiral fraction and low blue galaxy fractions indicators for maturity in a cluster core.

The other three cores show various properties indicating that they are less evolved, but no clear ranking appears. The Abell 901b core shows a particularly high S0 fraction with 44\% vs. $\sim 30$\% in the other cores and a correspondingly lower red spiral fraction of 8\% vs. $\sim 25$\% for the other cores, so it appears more evolved than A902 or the SW group. As Gray et al. (in preparation) point out, A901b has the highest X-ray luminosity of all four cores, so ram-pressure stripping by hot gas could well have been responsible for the final suppression of star formation and ensuing wiping-out of disk structure from former red spirals.

A902 in particular has the lowest spheroid fraction of all, with 50\% vs. $\sim 70$\% for the others, which means that A902 has the properties of a truly less dense region. Perhaps some line-of-sight projection has boosted its apparent density; \citet{Hey08} show from a weak-lensing analysis, that A902 is the lowest-mass structure among the four with a virial mass of $7\times 10^{13} M_\odot$ compared to $\sim 26\times 10^{13} M_\odot$ each for A901a and b. Further evidence for projection effects is discussed in Rhodes et al. (in preparation).

\section{Discussion}

\subsection{Red sequence contamination and the identification of star-forming samples}

It had been established by several authors, that red samples defined by CMR cuts do not only contain passively evolving spheroidal galaxies, but are contaminated with spiral galaxies both in clusters \citep{Po99} and in the field \citep{B04,Cas07}. There is also evidence for a large population of star-forming and at least moderately dusty galaxies in clusters (WGM05) or the field \citep{Gia05, Fr07,Ga08}, who report average fractions of 30-40\% in red samples, or fractions as a function of mass and environment: we see the most extreme environmental dependence at low mass, where fractions range from $<10$\% in cores to $>80$\% in the field, in excellent agreement with \citet{H08}. 

While \citet{Cas07} find red spirals in the field to be mostly edge-on galaxies, we emphasise that in dense environments inclination is not the dominant explanation for their colour: instead lower specific SFRs cause the red colour irrespective of a galaxy's orientation. The key is to look not at contamination fraction in red samples, but at the ratio of red to blue spirals instead: If red spirals were purely due to high inclination, random orientation statistics would suggest in a given mass bin, that the red fraction among spirals (or star-forming galaxies) should be independent of environment. In fact, the $z\approx 0.2$ field sample in WGM05 has a 1:8 ratio of dusty red to blue galaxies, indistinguishable from the red spiral to blue galaxy ratio at $z\approx 0.7$ reported by \citet{Cas07}. It is entirely plausible that one in eight spiral galaxies have inclinations high enough to extinguish most of the light from young disk stars due to the increased optical depth of dust projected on the disk. 

However, the red spiral fraction in our cluster sample exceeds dramatically the small geometric fraction, as it has {\it more red spirals than blue galaxies} at $\log M_*/M_\odot>10$. The majority of these show reduced star formation but do not show enhanced dust extinction of the overall stellar population. Their morphologies are also clearly not edge-on, but they have smoothed substructure as expected for a low-SFR system and seen in Fig.~\ref{CSseq}. Given that these galaxies are the major constituent of the star-forming population, we would like to include them when studying a star-forming sample. As a result, star-forming samples show different colour and star-formation properties than pure blue cloud samples.

WGM05 report an average equivalent width of [OII] emission in A901 dusty red galaxies of $\sim 4$\AA , compared to 17.5\AA \ in the blue cloud. [OII] line fluxes are an order of magnitude reduced in dusty red compared to blue galaxies. Hence, their [OII] emission would be undetectable in the cluster studies of \citet{VZG08} or in the VVDS field survey \citep{Fr07}. Their spectra otherwise appear similar to those of truly passive galaxies. It is thus no surprise that spectroscopic surveys have found red spirals {\it without optical signatures of star formation}. \citet{H08} point out that the SDSS fiber spectra at low redshift target a small nuclear region and fail to detect young stellar populations in many cases, a bias that enhances red-sequence contamination. In order to identify virtually all star-forming galaxies, they suggest to combine NUV data with an H$\alpha$-line sensitivity of 2\AA \ (equivalent width).

Finally we note, that \citet{Wei06} report a fraction of star-forming red galaxies (their 'intermediate types') among all galaxies in their SDSS sample ($M_r<-18$) of 20\%. Surprisingly, it appears to be entirely independent of the galaxy luminosity and its environment (halo mass, location in a galaxy group).

\subsection{The decline in star formation fractions and the average star formation rate}

A number of works have established two independent observations that are in apparent conflict: 

(1) The proximity of the Virgo cluster has permitted the discovery of a substantial population of spiral galaxies, which have truncated star-forming disks. The idea is that gas is gradually being stripped from galaxies, perhaps via ISM-ICM interaction, whereby the outer regions of a galaxy are swept first given their lower column densities \citep{KK04}, and the truncation radius shrinks with time. At intermediate redshifts there is further evidence that cluster spirals have more centrally-concentrated star formation than field galaxies \citep{BMA07}. Seeing such transition objects abundantly requires a long time scale for the SF-suppressing process.

(2) Most studies of SFR in star-forming galaxies report an independence of SFR on environment; they only find that the fraction of star-forming galaxies declines with density, thus requiring a fast transition \citep[e.g.][]{Ba04a}. The average equivalent width of the [OII] line in star-forming galaxies is found by \citet{Po08} at $z=[0.4,0.8]$ to be $\sim 15$\AA \ in the field, groups or clusters alike; in $z\approx 0.25$ clusters \citet{VZG08} find $\sim 16$\AA , and at $z\la 0.04$ \citet{H07} find 15\AA \ in bright $M_r<-20$ galaxies. Why this number should be constant across such a wide range in redshift is unclear, and may coincidentally arise from the mass-dependence of the S-SFR evolution with redshift and the redshift dependence of mass limits in the galaxy samples. 

Our results show that at $\log M_*/M_\odot=[10,11]$ the mean S-SFR of star-forming galaxies declines almost by 50\% from field to cores, but by much less in lower-mass galaxies. Thus, we have found a significant SFR decline with density, as a result of including dusty red galaxies in our star-forming sample and using more reliable SFR estimates. Had we restricted ourselves to blue galaxies alone, we would have seen only a 20\% decline with very large errors due to the smaller sample, instead of a clear 50\% decline. Also, our blue sample has [OII] properties similar to the literature samples above. 

The comparison between blue and star-forming samples helps to resolve the contradiction posed between observations of slow transformation in Virgo and apparently rapid truncation in more distant clusters: The inclusion of semi-passive spirals in the star-forming sample of cluster galaxies produces a decline of SFR in star-forming galaxies with density, which is consistent with slow quenching in clusters. In contrast, considerations of only blue-cloud galaxies require to invoke nearly instantaneous SF suppression given that their SFRs are apparently constant over density.

We note that \citet{H07} report a 20\% SFR decline across density in their faintest sample bin at $M_R=[-19,-18]$. Their brighter bins show no trends but are affected by an aperture bias, which they argue may hide evolution that may only play out in an extended star-forming disk outside the aperture. We believe that their judgement is correct; we have a smaller aperture bias in our data, given that our aperture has 4.25~kpc FWHM as opposed to the on average sub-kpc fiber aperture in \citet{H07}.

\subsection{The nature of red spirals or dusty red galaxies in clusters}

In $z\approx 0.4$ clusters \citet{Po99} reported seeing numerous {\it optically passive} spirals, i.e. spiral galaxies with an absorption-line spectrum characteristic of older stellar populations, and no [OII] emission or otherwise apparent star formation. They also found a population of {\it k+a galaxies} characterised by abrupt SF truncation and asked why these quiescent infall populations are so prevalent at $z\approx 0.4$, but apparently lacking at low $z$. In the Virgo cluster, star formation was still observed and spatially resolved in spirals, and it was later shown that the anemic spirals discovered by \citet{vdB76} were not the most abundant cluster-specific phenomenon \citep{KK04}; instead, the principal effect of the Virgo cluster on infalling spirals seemed to be a slow outside-in truncation of the star-forming disk.

In our young $z\approx 0.17$ cluster environment, we see again a very abundant infall population with similar optical properties as Poggianti's {\it optically passive} spirals, except that we have shown them to be on average only {\it semi-passive}: They still form stars at a substantial rate, as shown consistently from either our IR data or from our optical estimates of dust extinction and star formation. Their specific SFR, however, is on average lower by 0.6~dex in comparison to regular blue galaxies of the same mass. Given an S-SFR scatter in the blue cloud of $\sim 0.3$~dex \citep{Noe07,Sal07}, these semi-passive spirals are $\sim 2\sigma$-outliers from the regular star-forming sequence in field galaxy samples \citep[see also the discussion in][]{Schi07}. Altogether, we see spirals in the cluster to span the whole range from normally star-forming to virtually passive. We note that we do not see a star-bursting population in A901, and WGM05 found no evidence for an abundant post-starburst population either, although that was concluded from stacked spectra, and they could not reliably measure H$\delta$ equivalent widths in individual galaxies. 

WGM05 showed the optical spectral signatures of dusty red galaxies to be almost indistinguishable from passively evolving (old red) galaxies on the basis of high-S/N stacked spectra. Individual spectra were too noisy to find the weak [OII] lines and were otherwise dominated by the absorption-line spectrum of the older stars. It is now clear that the drop in star formation combined with an increase in the obscuration of star-forming regions, is sufficient to make the optical detection of star-formation signatures a challenge. 

But although the $U-V$ colours of red spirals and red spheroids are indistinguishable, this is only a coincidence of choosing two bands in which the effects of age and dust are ambiguous. Choosing a wider range of colour, it turns out that red spirals are redder than spheroids at the red end and bluer at the blue end of the spectrum, which was the critical differentiation used by WGM05. In Fig.~\ref{ssfrM} we saw how e.g. the use of a near-UV colour discriminates dusty red from old red galaxies. 

We thus suggest that the red, or optically passive, spirals in cluster environments discussed by various authors are the same phenomenon as our semi-passive dusty red galaxies. The differentiating criterion for separating them from canonical red-sequence galaxies was the spiral morphology in \citet{Po99} and despite similar $U-V$ colour the signature of dust and younger ages in the SED in WGM05. In some more star-forming and less obscured cases [OII] lines could be strong enough to appear above detection limits \citep{VZG08}; \citet{Wil08} also used MIR colours to trace star formation in red galaxies. In this paper, the morphology, the spiral-typical dust extinction, the brighter UV luminosity and the substantial IR fluxes all came together to form a consistent picture of semi-passive spirals.

Finally, we believe we can consider the SF-truncated Virgo spirals to be local counterparts to the dusty red galaxies as their properties match at least in global respects: \citet{KK04a} find the S-SFR in their Virgo spirals to be on average $2.5\times$ reduced compared to the field, with an overall range from 10-fold reduction to slight enhancement; this echoes the S-SFR properties of our dusty red galaxies. Also, the star formation in the Virgo spirals is on average more obscured, just as in dusty red galaxies again; in Virgo this is a consequence of the less obscured star formation in the outer disk being much reduced while the more obscured star formation in the central parts remains normal or is even enhanced.

Their appearance in large numbers suggests that spirals in many clusters undergo only slow quenching of their star formation, and that the time scale of morphological evolution must be longer than that of the spectral evolution. All this applies only in the mass regime of $\log M_*/M_\odot=[10,11]$, where the SFR decline from blue galaxies in the field to dusty red galaxies in the infall region is not accompanied by morphological changes. Morphology seems to change only further into the cluster alongside further SF decline that replaces dusty red galaxies by old red ones. At lower masses, in contrast, SFR decline and morphological changes appear more synchronised, and virtually no red spirals are found, except for a near-constant fraction of $<20$\% that may result from high-inclination spirals following random orientation statistics. This situation appears to be similar in the Virgo cluster where \citet{Bos08} find dwarf galaxies (defined by $\log L_H/L_\odot <9.6$) to be either star-forming or quiescent with no intermediate objects being found.

\subsection{Physical interpretation of the evidence}

Already Kennicutt (1983) observed that at fixed Hubble type Virgo galaxies were redder and less star-forming than their cousins in the field. \citet{BO85} confirmed the colour difference for further clusters. However, whether these trends are due to nature (hidden in the past) or due to nurture (change during cluster infall) could not be identified from snapshot pictures of individual galaxies or clusters, because a snapshot will show the effects of infall and cosmic evolution of the progenitors intertwined. If we can not see the histories of the individual galaxies, we have no means to assess whether the progenitors of present-day cluster galaxies looked like the present-day infall galaxies do. Only if this is the case, could we map the density axis onto a time axis and interpret the environmental trends as evolutionary tracks.

However, there is strong evidence, e.g., that higher-redshift clusters have fewer S0s and more spirals \citep{Dr97}, so for S0s the link is established that a build-up in time corresponds to a build-up in density. Higher-redshift cluster cores have similarities with lower-redshift cluster outskirts giving rise to a 'double downsizing' picture where star formation moves progressively to lower-mass objects as you move either forward in cosmic time or higher in density of the environment \citep{K04,Sm05,Tan05}.

Our results for A901/2 suggest, that empirically there are three mass domains with different behaviour along density:
\begin{itemize}
\item At high mass ($\log M_*/M_\odot>10.5$, but certainly at $>11$) only little change is observed with density, as most galaxies are red and spheroidal in all environments.
\item At intermediate mass ($\log M_*/M_\odot=[\sim10,\ga10.5]$) we see a sequence with density whereby blue spirals in the field are replaced by red spirals in the infall region, which are then replaced by S0s in the core. 
\item At low mass ($\log M_*/M_\odot\la10$) we see a sequence from blue field spirals and irregulars directly to red cluster spheroids.
\end{itemize}

Let us assume that the density axis can be mapped on a time arrow in the evolution of infalling galaxies, and the environmental trends are mostly due to nurture. An average galaxy falling in at 600~km/sec (the velocity dispersion of A901a/b dusty red galaxies listed, see WGM05) needs $\sim 2.5$~Gyr to fall from the virial radius to the core assuming a plunging orbit. There is thus no need to assume that they have already passed through the core. The results then say that star formation is suppressed at all densities and masses, and already between the field and the outermost infall region in our data, i.e. outside the virial radius of the cluster. This agrees with observations of galaxy pre-processing in infalling groups \citep[e.g.][]{Kod01,Wil05}. Morphological transformation appears in step with the SFR decline in low-mass galaxies, but is delayed for intermediate-mass galaxies. The delay gives rise to spiral galaxies in the infall region that appear red due to a combination of moderate extinction and reduced star formation. They appear broadly less clumpy and have less well-defined spiral arms than their blue cousins in the field, and they are the only progenitor candidates of similar-mass S0 galaxies. 

The steeper S-SFR decline for massive galaxies demonstrates that they transform slowly and remain visible in transformation for long. In contrast, the near-absence of a trend in mean S-SFR at lower mass requires more abrupt conversions that leave little chance to observe transforming objects, while giving rise to the steep trends in the SF/non-SF composition. We roughly find (with large errors) that the red fraction of spirals is increased in the infall region by perhaps 10\% over the field value at a mass of $\log M_*/M_\odot\approx 9.75$ and by maybe 50\% at $\log M_*/M_\odot\approx 10.5$. At equal infall velocity this would imply $5\times$ longer quenching time-scales at $5\times$ higher mass or $t_{\rm trans}\propto M_*$. The time-scale for ram-pressure stripping though is $t_{\rm RPS}\propto r_{\rm gal} \sqrt{\rho_{\rm ISM}} v_{\rm rel}^{-1} \propto \sqrt{\rho_{\rm ISM} M_*}$ assuming constant surface mass density of disk galaxies \citep{Bar05} and equal infall velocity. While we do not have the data to be very quantitative here, we argue that at least the quenching time-scale is unlikely to decline with mass.

\citet{Bos08} model the evolution of dwarf galaxies in Virgo (with $\log L_H/L_\odot<9.6$) assuming interaction with the intracluster medium, and derive SF quenching time scales of $<150$~Myr. Such a rapid SF truncation would be consistent with the absence of a noticeable transition population. Also, \citet{Po04} see k+a galaxies in the Coma cluster only fainter than $M_V=-18.5$: the k+a signature requires a sharp truncation in star formation, which argues in favour of a faster shutdown of star formation at lower mass. In higher-mass spirals, gas stripping by hot ICM would first remove gas from the outer disks where the gas column densities are lower, and the slow truncation of star-forming disks seen by \citet{KK04} would leave morphology largely intact over a slow SF quenching process, and would hence not lead to a k+a signature. 

It then may remain as puzzling that \citet{Po99} found many k+a spirals of normal mass in clusters at $z\approx 0.4$. However, the transformation time scale must depend not only on galaxy mass but also on the ICM properties. Already \citet{Dr80} found higher S0 fractions with increasing X-ray luminosity of the clusters, although the presence of a residual relation at fixed cluster mass is unclear \citep[see discussion in][]{Po05}. However, the basic effect has been studied by comparing two $z\approx 0.5$ clusters with very different ICM properties representing a low-mass (Cl0024) and high-mass (MS0451) prototype in cluster evolution \citep{Mo07}. Low-mass clusters are irregular with several sub-clumps, perhaps young and still assembling, and have little volume occupied by hot ICM (e.g. Virgo and A901/2); higher-mass clusters are more regular in structure and have a large volume occupied by hot ICM (e.g. Coma). The galaxies in the two clusters compared by \citet{Mo07} are on two different tracks of S-SFR-vs.-age as seen from [OII]-vs.-$D_{4000}$, showing that ram pressure from the well-developed ICM in MS0451 shuts down star formation rapidly even in larger galaxies, while the less developed ICM in Cl0024 means the transition is a lot slower and may even be driven predominantly by factors other than stripping by the ICM, so that many transitioning red spirals with low S-SFR are seen. In line with this picture, we already see variation between our four cluster cores: the most X-ray-luminous core A901b has not only the most prominent ICM but also the lowest red spiral and the highest S0 fraction. 

The discussion whether the observed S0 galaxies have evolved from spiral progenitors has continued for three decades now \citep{Dr80,Mo07}. We suggest to tackle the question with a mass-resolved approach, motivated by the strong trends seen at intermediate mass in our cluster. Several authors \citep[e.g][]{Bur05} have argued that present-day S0s are too luminous to be made from present-day infalling spirals. In A901/2 we admittedly find S0 galaxies, but no infalling spirals at $\log M_*/M_\odot>11$; we suggest that these would have formed in the more distant past of this cluster from more massive progenitors that we do not observe in the present snapshot due to downsizing. We have no evidence against this picture, and note the absence of massive star-forming galaxies that could further enrich the massive S0 sample in A901/2 in the near future. In contrast, at $\log M_*/M_\odot=[10,10.5]$ we find a very strong trend of red spirals disappearing towards the core and S0s appearing in comparable numbers; the S0 concentration towards the cores of A901/2 is markedly different from the flat S0 fraction within a virial radius seen in ensembles of more evolved clusters \citep{Dr80,Dr97}. This is consistent with A901/2 being a young clusters with weaker ICM, and it suggests that A901/2 has much progenitor material to form more intermediate-mass S0s over the next few Gyrs to fill the virial volume more evenly.

A further obstacle in the understanding of S0 formation is the difference in the structural properties of S0s compared to their progenitor spirals: S0s are claimed to have larger concentrations or bulges and thicker disks than their progenitors, although recent works claim a smaller degree of difference \citep{Lau07,Wz08}. We are not aware of a mass-resolved study of progenitors and S0s that evaluates the effects of gradual gas stripping and continuing centrally concentrated star formation. \citet{Bl05} studied residual morphological trends in the SDSS data after eliminating the primary trend of the dominant colour-density relation and found only a weak trend towards higher concentration with density, and applying only to more luminous galaxies. This would be entirely consistent with a picture where low-mass galaxies experience more abrupt SF shutoff in dense environments, while high-mass galaxies continue forming stars in their inner regions. However, \cite{K04} find no change in concentration with density at $\log M_*/M_\odot>10.5$, and only a weak trend below.

\section{Conclusions}

We have investigated the properties of galaxies, especially those of optically passive spirals, in a cluster complex at redshift $\sim 0.17$, which consists of four cores (A901a, A901b, A902 and SW group). We have used the STAGES dataset published by \citet{G08}, exploiting the restframe near-UV-optical SEDs, IR 24$\mu$m data and HST morphologies. The cluster sample is defined on the basis of COMBO-17 redshifts with an rms precision of $\sigma_{cz}\approx 2000$~km/sec (at $M_V<-19$). We draw the following conclusions:

\begin{enumerate}[1.]
\item Optically passive spirals in clusters, dusty red galaxies in A901/2 and Virgo spirals with truncated star-formation disks \citep{Po99,WGM05,KK04} appear to be basically the same phenomenon. However, these objects are not truly passive galaxies despite their integrated optical appearance from large distances.
\item They form stars at a substantial rate but reduced in comparison to field spirals (on average $4\times$ lower SFR at fixed mass in our sample). However, their star formation is more obscured and its optical signatures are weak.
\item They appear predominantly in the mass range $\log M_*/M_\odot=[10,11]$ where they constitute over half of the star-forming galaxies in the cluster, and thus form an important transition population.
\item We find that the mean S-SFR of star-forming galaxies in the cluster is clearly lower than in the field, in contrast to the S-SFR properties of blue galaxies alone, which appear unchanged. The dusty red spirals are thus a vital ingredient for understanding the overall picture of SF quenching in clusters.
\item At lower mass the dusty red sample consists mostly of edge-on spirals, which are a small fraction of all spirals with no environmental dependence. There is limited room for a cluster-specific contribution.
\item Physically, it seems that star formation quenching is fast in low-mass galaxies and accompanied by  morphological change; hence, no cluster-specific red spiral phenomenon is observed at $\log M_*/M_\odot<10$. At larger masses, SF quenching is a slower process and strong morphological transformation is even more delayed, thus giving rise to abundant red spirals.
\item The currently observed red spirals are expected to turn into S0s with time when their star formation is terminated by hot ICM, although they may end up with different detailed properties from those S0s that are already in place.
\end{enumerate}

\section*{acknowledgements}
We thank the anonymous referee for a very careful reading of the paper and very helpful suggestions improving the manuscript. We also thank Richard S. Ellis, Benjamin D. Johnson, Tommaso Treu for comments on this work. Support for STAGES was provided by NASA through GO-10395 from STScI operated by AURA under NAS5-26555.  MEG and CW were supported by STFC Advanced Fellowships. CH acknowledges the support of a European Commission Programme 6th framework Marie Cure Outgoing International Fellowship under contract MOIF-CT- 2006-21891. CYP was supported by the NRC-HIA Plaskett Fellowship, and the STScI Institute/Giacconi Fellowship. EFB, AG and KJ are greatful for support from the DFG's Emmy Noether Programme of the Deutsche Forschungsgemeinschaft, AB by DLR grant no. 50 OR 0404, DHM by NASA under LTSA Grant NAG5-13102, MB and EvK by the Austrian Science Foundation FWF under grant P18416, SFS by the Spanish MEC grants AYA2005-09413-C02-02 and the PAI of the Junta de Andaluca as research group FQM322, SJ by NASA under LTSA Grant NAG5-13063, GO-10861, G0-11082, and by NSF under AST-0607748.


\begin{thebibliography}{}
\bibitem[\protect\citeauthoryear{Balogh et al.}{2000}]{Ba00} 
  Balogh, M. L., Navarro, J. F. \& Morris, S. L., 2000, ApJ, 540, 113
\bibitem[\protect\citeauthoryear{Balogh et al.}{2004a}]{Ba04a} 
  Balogh, M. L., Eke, V., Miller, C., Lewis, I. et al, 2004, MNRAS, 348, 1355
\bibitem[\protect\citeauthoryear{Balogh et al.}{2004b}]{Ba04b} 
  Balogh, M. L., Baldry, I. K., Nichol, R., Miller, C., Bower, R., Glazebrook, K., 2004, ApJ, 615, L101
\bibitem[\protect\citeauthoryear{Bamford et al.}{2007}]{BMA07}
  Bamford, S. P.,Milvang-Jensen, B. \& Arag{\'o}n-Salamanca, A., 2007, MNRAS, 378, L6
\bibitem[\protect\citeauthoryear{Bamford et al.}{2008}]{Bam08}
  Bamford, S. P., Nichol, R. et al, 2008, MNRAS, submitted (astro-ph/0805.2612)
\bibitem[\protect\citeauthoryear{Barden et al.}{2005}]{Bar05}
  Barden, M., Rix, H.-W., Somerville, R. S., et al., 2005, ApJ, 635, 959
\bibitem[\protect\citeauthoryear{Barnes}{1992}]{Ba92}
  Barnes, J. E., 1992, ApJ, 393, 484
\bibitem[\protect\citeauthoryear{Bekki}{1999}]{Be99}
  Bekki, K., 1999, ApJ, 510, L15
\bibitem[\protect\citeauthoryear{Bell et al.}{2004}]{B04}
  Bell, E. F., McIntosh, D. H., Barden, M., Wolf, C. et al., 2004, ApJ, 600, L11
\bibitem[\protect\citeauthoryear{Bell et al.}{2007}]{B07}
  Bell, E. F., Zheng, X. Z., Papovich, C., Borch, A., Wolf, C. \& Meisenheimer, K., 2007, ApJ, 663, 834
\bibitem[\protect\citeauthoryear{Benson et al.}{2003}]{Be03}
  Benson, A. J., Bower, R. G., Frenk, C. S., Lacey, C. G., Baugh, C. M., Cole, S., 2003, ApJ, 599, 38
\bibitem[\protect\citeauthoryear{Blanton et al.}{2005}]{Bl05}
  Blanton, M. R., Eisenstein, D., Hogg, D. W., Schlegel, D. J., Brinkmann, J., 2005, ApJ, 629, 143
\bibitem[\protect\citeauthoryear{Boselli et al.}{2008}]{Bos08}
  Boselli, A., Boissier, S., Cortese, L. \& Gavazzi, G., 2008, ApJ, 674, 742
\bibitem[\protect\citeauthoryear{Butcher \& Oemler}{1985}]{BO85}
  Butcher, H. \& Oemler, A., Jr., 1985, ApJS, 57, 665
\bibitem[\protect\citeauthoryear{Bundy et al.}{2006}]{Bu06}
  Bundy, K., Ellis, R. S., Conselice, C. J., Taylor, J. E. et al, 2006, ApJ, 651, 120
\bibitem[\protect\citeauthoryear{Burstein et al.}{2005}]{Bur05}
  Burstein, D., Ho, L. C., Huchra, J. P. \& Macri, L. M., 2005, ApJ, 621, 246
\bibitem[\protect\citeauthoryear{Calzetti}{2001}]{Cal01}
  Calzetti, D., 2001, PASP, 113, 1449
\bibitem[\protect\citeauthoryear{Calzetti et al.}{1994}]{Ca94}
  Calzetti, D., Kinney, A. \& Storchi-Bergmann, T., 1994, ApJ, 429, 582
\bibitem[\protect\citeauthoryear{Cassata et al.}{2007}]{Cas07}
  Cassata, P., Guzzo, L., Frnceschine, A., Scoville, N., et al., 2007, ApJS, 172, 270
\bibitem[\protect\citeauthoryear{Cooper et al.}{2008}]{Co08}
  Cooper, M. C., Newman, J. A., Weiner, B. J., Yan, R., Willmer, C. N. A. et al., 2008, MNRAS, 383, 1058
\bibitem[\protect\citeauthoryear{Croton et al.}{2006}]{Cr06}
  Croton, D. J., Springel, V., White, S. D. M. et al, 2006, MNRAS, 365, 11
\bibitem[\protect\citeauthoryear{Christlein \& Zabludoff}{2004}]{CZ04}
  Christlein, D. \& Zabludoff, A. I., 2004, ApJ, 616, 192
\bibitem[\protect\citeauthoryear{Christlein \& Zabludoff}{2005}]{CZ05}
  Christlein, D. \& Zabludoff, A. I., 2005, ApJ, 621, 201
\bibitem[\protect\citeauthoryear{Dressler}{1980}]{Dr80}
  Dressler, A., 1980, ApJ, 236, 351
\bibitem[\protect\citeauthoryear{Dressler}{1997}]{Dr97}
  Dressler, A., Oemler, A., Couch, W. J., Smail, I. et al. 1997, ApJ, 490, 577 
\bibitem[\protect\citeauthoryear{Elbaz et al.}{2007}]{El07}
  Elbaz, D., Daddi, E., LeBorgne, E., Dickinson, M. et al, 2007, A\&A, 468, 33
\bibitem[\protect\citeauthoryear{Franzetti et al.}{2007}]{Fr07}
  Franzetti, P., Scodeggio, M., Garilli, B., Vergani, D. et al., 2007, A\&A, 465, 711
\bibitem[\protect\citeauthoryear{Gallazzi et al.}{2008}]{Ga08}
  Gallazzi, A., Bell, E. F., Wolf, C. et al, 2008, ApJ, in press
\bibitem[\protect\citeauthoryear{Giallongo et al.}{2005}]{Gia05}
  Giallongo, E., Salimbeni, S., Menci, N., Zamorani, G. et al. 2005, ApJ, 622, 116
\bibitem[\protect\citeauthoryear{Gilmour et al.}{2007}]{Gil07}
  Gilmour, R., Gray, M. E., Almaini, O., Best, P., Wolf, C., Meisenheimer, K., Papovich, C. 
  \& Bell, E. F., 2007, MNRAS, 380, 1467
\bibitem[\protect\citeauthoryear{Gomez et al.}{2003}]{Go03}
  Gomez, P. L., Nichol, R. C., Miller, C. J. et al, 2003, ApJ, 584, 210
\bibitem[\protect\citeauthoryear{Gray et al.}{2008}]{G08}
  Gray, M. E., Wolf, C., Barden, M., Peng, C. Y. et al, 2008, MNRAS, in press (this issue)
\bibitem[\protect\citeauthoryear{Gunn \& Gott}{1972}]{GG72}
  Gunn, J. E. \& Gott, J. R., 1972, ApJ, 176, 1
\bibitem[\protect\citeauthoryear{Guzzo et al.}{2007}]{Gu07}
  Guzzo, L., Cassata, P., Finoguenov, A., Massey, R., Scoville, N. Z. et al, 2007, ApJS, 172, 254
\bibitem[\protect\citeauthoryear{Haines et al.}{2007}]{H07}
  Haines, C. P., Fargiulo, A., La Barbera, F., Mercurio, A., Merluzzi, P. \& Busarello, G.,
  2007, MNRAS, 381, 7
\bibitem[\protect\citeauthoryear{Haines et al.}{2008}]{H08}
  Haines, C. P., Gargiulo, A. \& Merluzzi, P., 2008, MNRAS, 385, 1201
\bibitem[\protect\citeauthoryear{Heiderman et al.}{2008}]{Hei08}
  Heiderman, A., Jogee, S., Marinova, I., van Kampen, E. et al., 2008, ApJ, submitted
\bibitem[\protect\citeauthoryear{Heymans et al.}{2008}]{Hey08}
  Heymans, C., Gray, M. E., Peng, C. Y., van Waerbeke, L. et al., 2008, MNRAS, 385, 1431
\bibitem[\protect\citeauthoryear{Kauffmann et al.}{2003}]{Kau03}
  Kauffmann, G., Heckman, T. M., White, S. D. M. et al., 2003, MNRAS, 341, 54
\bibitem[\protect\citeauthoryear{Kauffmann et al.}{2004}]{K04}
  Kauffmann, G., White, S. D. M., Heckman, T. M. et al., 2004, MNRAS, 353, 713
\bibitem[\protect\citeauthoryear{Koopman \& Kenney}{1998}]{KK98}
  Koopman, R. A. \& Kenney, J. D. P., 1998, ApJ, 497, L75
\bibitem[\protect\citeauthoryear{Koopman \& Kenney}{2004a}]{KK04a}
  Koopman, R. A. \& Kenney, J. D. P., 2004, ApJ, 613, 851
\bibitem[\protect\citeauthoryear{Koopman \& Kenney}{2004b}]{KK04}
  Koopman, R. A. \& Kenney, J. D. P., 2004, ApJ, 613, 866
\bibitem[\protect\citeauthoryear{Kenney et al.}{2008}]{Ke08}
  Kenney, J. D. P., Wong, O. I. et al., 2008, submitted (astro-ph/0803.2532)
\bibitem[\protect\citeauthoryear{Kennicutt}{1983}]{K83}
  Kennicutt, R. C., 1983, AJ, 88, 483
\bibitem[\protect\citeauthoryear{Kennicutt}{1998}]{K98}
  Kennicutt, R. C., 1998, ARA\&A, 36, 189
\bibitem[\protect\citeauthoryear{Khochfar \& Ostriker}{2008}]{KO08}
  Khochfar, S. \& Ostriker, J., 2008, ApJ, 680, 54
\bibitem[\protect\citeauthoryear{Kodama et al.}{2001}]{Kod01}
  Kodama, T., Smail, I., Nakata, F., Okamura, S. \& Bower, R. G., 2001, ApJ, 562, L9
\bibitem[\protect\citeauthoryear{Lane et al.}{2007}]{L07}
  Lane, K. P., Gray, M. E., Arag{\'o}n-Salamanca, A., Wolf, C. \& Meisenheimer, 2007, MNRAS, 378, 716
\bibitem[\protect\citeauthoryear{Larson et al.}{1980}]{La80}
  Larson, R. B., Tinsley, B. M. \& Caldwell, C. N., 1980, ApJ, 237, 692
\bibitem[\protect\citeauthoryear{Laurikainen et al.}{2007}]{Lau07}
  Laurikainen, E., Salo, H., Buta, R. \& Knapen, J. H., 2007, MNRAS, 381, 401
\bibitem[\protect\citeauthoryear{Lewis et al.}{2002}]{Le02}
  Lewis, I., Balogh, M., De Propris, R., Couch, W. et al., 2002, MNRAS, 334, 673
\bibitem[\protect\citeauthoryear{Marinova et al.}{2008}]{Ma08}
  Marinova, I., Jogee, S., Heiderman, A. et al., 2008, ApJ, submitted
\bibitem[\protect\citeauthoryear{Mehlert et al.}{2003}]{Me03}
  Mehlert, D., Thomas, D., et al., 2003, A\&A, 407, 423
\bibitem[\protect\citeauthoryear{Moore et al.}{1996}]{Mo96}
  Moore, B., Katz, N., Lake, G., Dressler, A. \& Oemler, A., 1996, Nature, 379, 613
\bibitem[\protect\citeauthoryear{Moran et al.}{2007}]{Mo07}
  Moran, S. M., Ellis, R. S., Treu, T. et al, 2007, ApJ, 671, 1503
\bibitem[\protect\citeauthoryear{Moss}{2006}]{Mo06}
  Moss, C., 2006, MNRAS, 373, 167
\bibitem[\protect\citeauthoryear{Moss \& Whittle}{2000}]{MW00}
  Moss, C. \& Whittle, M., 2000, MNRAS, 317, 667
\bibitem[\protect\citeauthoryear{Noeske et al.}{2007}]{Noe07}
  Noeske, K. G., Weiner, B. J., Faber, S. M., Papovich, C. et al., 2007, ApJ, 660, L43
\bibitem[\protect\citeauthoryear{Park et al.}{2007}]{P07}
  Park, C., Choi, Y.-Y., Vogeley, M. S., Gott, R. (III) \& Blanton, M. R., 2007, ApJ, 658, 898
\bibitem[\protect\citeauthoryear{Park et al.}{2008}]{P08}
  Park, C., Gott, R. \& Choi, Y.-Y., 2008, ApJ, 674, 784
\bibitem[\protect\citeauthoryear{Pei}{1992}]{P92}
  Pei, Y. C., 1992, ApJ, 395, 130
\bibitem[\protect\citeauthoryear{Peng et al.}{2002}]{Pe02}
  Peng, C. Y., Ho, L. C., Impey, C. D. \& Rix, H.-W., 2002, AJ, 124, 266
\bibitem[\protect\citeauthoryear{Poggianti et al.}{1999}]{Po99}
  Poggianti, B. M., Smail, I., Dressler, A., Couch, W. J. et al., 1999, ApJ, 518, 576
\bibitem[\protect\citeauthoryear{Poggianti et al.}{2004}]{Po04}
  Poggianti, B. M., Bridges, T. J., Komiyama, Y., Yagi, M. et al., 2004, ApJ, 601, 197
\bibitem[\protect\citeauthoryear{Poggianti et al.}{2008}]{Po08}
  Poggianti, B. M., Desai, V., Finn, R. et al., 2008, ApJ, 684, 888
\bibitem[\protect\citeauthoryear{Postman et al.}{2005}]{Po05}
  Postman, M., Franx, M., Cross, N. J. G., Holden, B. et al., 2005, ApJ, 623, 721
\bibitem[\protect\citeauthoryear{Rines \& Geller}{2008}]{RG08}
  Rines, K. \& Geller, M. J., 2008, AJ, 135, 1837
\bibitem[\protect\citeauthoryear{Sabatini et al.}{2003}]{Sab03}
  Sabatini, S., Davies, J., Scaramella, R., Smith, R. et al., 2003, MNRAS, 341, 981
\bibitem[\protect\citeauthoryear{Salim et al.}{2007}]{Sal07}
  Salim, S., Rich, R. M., Charlot, S. et al., 2007, ApJS, 173, 267 
\bibitem[\protect\citeauthoryear{Salpeter}{1955}]{S55}
  Salpeter, E. E. 1955, ApJ, 121, 161
\bibitem[\protect\citeauthoryear{Sandage et al.}{1985}]{SBT85}
  Sandage, A., Binggeli, B. \& Tammann, G. A., 1985, AJ, 90, 1759
\bibitem[\protect\citeauthoryear{Schawinski et al.}{2007}]{Sch07}
  Schawinksi, K., Thomas, D., et al., 2007, MNRAS, 382, 1415
\bibitem[\protect\citeauthoryear{Schiminovich et al.}{2007}]{Schi07}
  Schiminovich, D., Wyder, T. K., Martin, D. C., et al., 2007, ApJS, 173, 315
\bibitem[\protect\citeauthoryear{Smith et al.}{2005}]{Sm05}
  Smith, G. P., Treu, T., Ellis, R. S., Moran, S. M. \& Dressler, A., 2005, ApJ, 620, 78
\bibitem[\protect\citeauthoryear{Tanaka et al.}{2005}]{Tan05}
  Tanaka, M., Kodama, T. et al., 2005, MNRAS, 362, 268 
\bibitem[\protect\citeauthoryear{Thomas et al.}{2005}]{Th05}
  Thomas, D., Maraston, C., Bender, R. \& Mendes de Oliveira, C., 2005, ApJ, 621, 673
\bibitem[\protect\citeauthoryear{Trager et al.}{2008}]{Tr08} 
  Trager, S. C., Faber, S. M. \& Dressler, A., 2008, MNRAS, 386, 715
\bibitem[\protect\citeauthoryear{Tremonti et al.}{2007}]{Tr07} 
  Tremonti, C. A., Moustakas, J. \& Diamond-Stanic, A. M., 2007, ApJ, 663, L77
\bibitem[\protect\citeauthoryear{van den Bergh}{1976}]{vdB76}
  van den Bergh, S., 1976, ApJ, 206, 883
\bibitem[\protect\citeauthoryear{van der Wel}{2008}]{vdW08}
  van der Wel, A., 2008, ApJ, 675, L13
\bibitem[\protect\citeauthoryear{Verdugo et al.}{2008}]{VZG08}
  Verdugo, M., Ziegler, B. L. \& Gerken, B., 2008, A\&A, 486, 9
\bibitem[\protect\citeauthoryear{Wilman et al.}{2005}]{Wil05}
  Wilman, D. J., Balogh, M. L. et al., 2005, MNRAS, 358, 71
\bibitem[\protect\citeauthoryear{Wilman et al.}{2008}]{Wil08}
  Wilman, D. J., Pierini, D., Tyler, K. et al., 2008, ApJ, 680, 1009
\bibitem[\protect\citeauthoryear{Weinmann et al.}{2006}]{Wei06}
  Weinmann, S. M., van den Bosch, F. C., Yang, X. \& Mo, H. J., 2006, MNRAS, 366, 2 
\bibitem[\protect\citeauthoryear{Weinzirl et al.}{2008}]{Wz08}
  Weinzirl, T., Jogee, S., Khochfar, S., Burkert, A. \& Kormendy, J., 2008, ApJ, submitted
  (astro-ph/0807.0040)
\bibitem[\protect\citeauthoryear{Wolf et al.}{2003}]{W03}
  Wolf, C., Meisenheimer, K., et al., 2003, A\&A, 401, 73
\bibitem[\protect\citeauthoryear{Wolf et al.}{2004}]{W04}
  Wolf, C., Meisenheimer, K., Kleinheinrich, M. et al., 2004, A\&A, 421, 913
\bibitem[\protect\citeauthoryear{Wolf et al.}{2005}]{WGM05}
  Wolf, C., Gray, M., \& Meisenheimer, 2005, A\&A, 443, 435
\bibitem[\protect\citeauthoryear{Wolf et al.}{2007}]{W07} 
  Wolf, C., Gray, M. E., Arag{\'o}n-Salamanca, A., Lane, K. P. \& Meisenheimer, 2007, MNRAS, 376, L1
\bibitem[\protect\citeauthoryear{Wolf et al.}{2008}]{W08}
  Wolf, C., Hildebrandt, H., Taylor, E. N. \& Meisenheimer, K., 2008, A\&A, in press
\bibitem[\protect\citeauthoryear{Xu et al.}{2004}]{Xu04}
  Xu, C. K., Sun, Y. C. \& He, X. T., 2004, ApJ, 603, L73
\end{thebibliography}
\end{document}